\documentclass[12pt,preprint]{aastex}
\usepackage{epsfig}
\usepackage{subfigure}
\usepackage{amssymb}

\renewenvironment{figure}{\begin{figure*} }{\end{figure*}}

\newcommand{\bu}{{\bf u}}

\newcommand{\bg}{{\bf g}}

\newcommand{\be}{{\bf e}}
\newcommand{\grad}{{\mathbf \nabla}}
\renewcommand{\div}{{\mathbf \nabla} \cdot}

\newcommand{\ephi}{\be_\phi}

\newcommand{\ez}{\be_z}
\newcommand{\dd}{{\rm d}}

\newcommand{\Enu}{E_\nu}

\begin{document}

\title{Gyroscopic pumping of large-scale flows in stellar interiors, 
and application to Lithium Dip stars}

\author{P. Garaud \& P. Bodenheimer} 

\affil{Department of Applied Mathematics and Statistics, Baskin School of
 Engineering,  and UCO/Lick Observatory, Department of Astronomy and 
Astrophysics, University of California Santa Cruz, 1156 High Street, CA 95064 
Santa Cruz, USA}

\maketitle

\section*{Abstract}
The maintenance of large-scale differential rotation 
in stellar convective regions by rotationally influenced
convective stresses also drives large-scale meridional flows
by angular--momentum conservation. This process is an 
example of ``gyroscopic pumping'', and has recently been 
studied in detail in the solar context. An important question 
concerns the extent to which these gyroscopically pumped 
meridional flows penetrate into nearby 
stably stratified (radiative) regions, since they could 
potentially be an important source of non-local mixing. 
Here we present an extensive study of the gyroscopic pumping mechanism,
using a combination of analytical calculations and
numerical simulations both in Cartesian geometry and in
spherical geometry. The various methods, when compared with 
one another, provide physical insight into the process itself, as well
as increasingly sophisticated means of estimating the gyroscopic pumping rate. 
As an example of application, we investigate the effects of this 
large-scale mixing process on the surface abundances of the light elements
Li and Be for stars in the mass range 1.3--1.5$M_\odot$ 
(so-called ``Li-dip stars''). We 
find that gyroscopic pumping is a very efficient mechanism 
for circulating material between the surface and the deep interior, 
so much in fact that it over-estimates Li and Be depletion 
by orders of magnitude for stars on the hot side of the dip.
However, when the diffusion of chemical species back into the 
surface convection zone is taken into account, a good fit with 
observed surface abundances of Li and Be as a function of stellar mass 
in the Hyades cluster can be found for reasonable choices of model 
parameters.


\keywords{hydrodynamics --- method:numerical --- method:analytical --- stars:abundances --- stars:interiors --- stars: rotation}

\section{Introduction}
\label{sec:intro}

\subsection{Gyroscopic pumping}
\label{sec:introgyro}

Recently, Garaud \& Acevedo-Arreguin (2009) (GAA09 hereafter) presented 
a preliminary analysis of a ``new'' mechanism for rotational mixing 
in stellar interiors and conjectured on its potential 
role in the depletion of Lithium in young Main Sequence stars. 
This mechanism, called ``gyroscopic pumping'', 
was originally studied in the context of Earth's 
atmospheric dynamics by Haynes et al. (1991) and later discussed in the 
astrophysical context by Gough \& McIntyre (1998) and McIntyre (2007)
who argued that it plays an important role in the dynamics of the solar 
interior.
Loosely speaking, gyroscopic pumping occurs in any rotating fluid
in the presence of additional stresses or forces which perpetually 
accelerate or decelerate the flows in the azimuthal direction. 
By conservation of angular momentum, the accelerated parcels of fluid
move away from the rotation axis while the decelerated ones move
toward the rotation axis, thus generating meridional fluid motion.

\begin{figure}[h]
\centerline{\epsfig{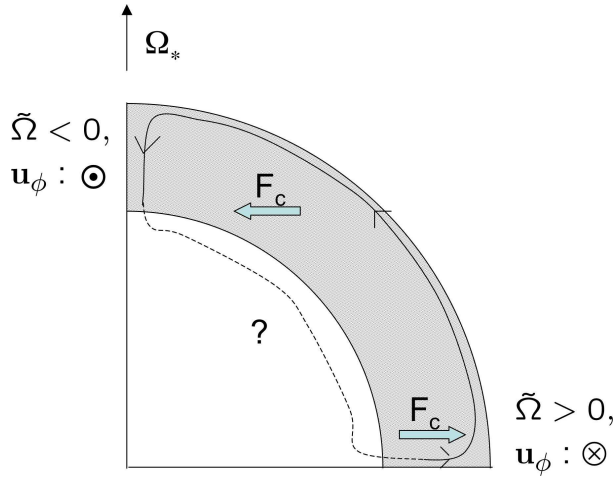}}
\caption{Illustration of gyroscopic pumping in a star with an outer 
convection zone (shaded area). Slower rotating regions near the poles 
pump the fluid 
towards the rotation axis through the action of the Coriolis force 
(${\bf F}_{\rm c}$) while the rapidly rotating region near the equator 
pump the fluid away 
from the rotation axis. A large-scale circulation is thus driven, 
poleward near the surface. The position of the return flow depends 
on the radiative zone physics (see GAA09 and \S\ref{sec:intromix} 
for detail). }
\label{fig:gyro} 
\end{figure}

Stars with outer convective regions often exhibit 
a significant amount of surface differential rotation 
(e.g. Collier-Cameron 2007; Reiners 2007). 
This differential
rotation is thought to be maintained by anisotropic and spatially 
varying Reynolds stresses (see R\"udiger 1989, for example), 
which tend to continually
accelerate the equatorial regions and decelerate the poles in a manner
most remarkably observed in the solar convection zone (Schou et al. 1998). 
If one assumes that the star is close to dynamical equilibrium, 
its mean rotation rate $\Omega_{\star}$ lies 
in between the polar and equatorial rotation rates. 
When viewed in a frame rotating with angular velocity $\Omega_{\star}$, 
the differential rotation of the star's convective zone forms
a large-scale azimuthal flow pattern, typically prograde in the 
equatorial region and retrograde in the polar regions. The aforementioned 
 ``gyroscopic pumping'' can then be viewed in two equivalent ways. 
As described earlier the 
constant acceleration of the equatorial regions is a local source of 
angular-momentum to the fluid, which by angular-momentum conservation must
move outward from the rotation axis. Similarly, fluid in the polar 
regions must move toward the rotation axis. Alternatively, one may simply 
note that the Coriolis force associated with the azimuthal flows described
above pushes the fluid away from the rotation axis in prograde regions, and
toward the rotation axis in retrograde regions. 
As illustrated in Figure \ref{fig:gyro}, 
the process naturally drives large-scale meridional flows throughout
the outer convection region, with an upwelling near the equator, 
a poleward velocity near the surface and downwelling near the poles. 
The polar downwelling does not need to stop at the radiative--convective
interface, and could in principle cause significant non-local 
mixing between the outer convection zone and the regions below. 

A similar gyroscopic pumping process is likely to drive fluid motion from  
within an inner convective zone as well. While the internal rotation profile 
of stars with convective cores has never been observed, one
can readily expect some degree of differential rotation
since the turbulent stresses associated with the rotationally 
constrained convective motions are likely to act in a similar fashion 
to those of the outer convective region. 
In that case again, there is no a-priori
reason for the flows thus generated to stop at the interface with the overlying
radiative zone, and one may wonder how much mixing they induce
in the star.

\subsection{Mixing in radiative regions induced by gyroscopic pumping}
\label{sec:intromix}

The first quantitative 
study of gyroscopic pumping in the context of stellar interiors
was recently presented by GAA09. They focused on the solar case, i.e. a star
with an outer convective region and an inner radiative region, and worked
in the Boussinesq approximation (arguing that the solar radiative zone 
does not span too many pressure and density scaleheights). 
They showed that the fate of gyroscopically 
pumped flows -- how much overall mixing they induce beyond the 
convective zone -- depends equally on the thermal stratification and on the 
dynamical properties of the nearby radiative region. 

In accordance with the earlier results of Garaud \& Brummell (2008), they 
found that in stratified, rotating stars in quasi-steady dynamical balance, 
the effect of stratification on large-scale meridional flows is principally
controlled by the quantity 
\begin{equation}
\sigma = \sqrt{\rm Pr} \frac{N}{\Omega_\star} \mbox{  , }
\label{eq:sigmadef}
\end{equation}
where Pr $=\nu/\kappa$ is the Prandtl number (where $\nu$ is the local 
viscosity, and $\kappa$ is the local thermal diffusivity), $N$ is the local
Brunt-V\"ais\"al\"a -- or buoyancy -- frequency and $\Omega_\star$ is the 
mean stellar 
rotation rate. When $\sigma$ is large, the effect of thermal stratification 
is strongly felt by the gyroscopically pumped meridional flows, which are 
exponentially damped 
away from the radiative--convective interface on the lengthscale 
$R_\star / \sigma$. Correspondingly, if  $\sigma$ is small the 
flows can {\it in principle} 
penetrate much more deeply into the radiative interior, and cause significant
large-scale mixing.

\begin{figure}[h]
\centerline{\epsfig{file=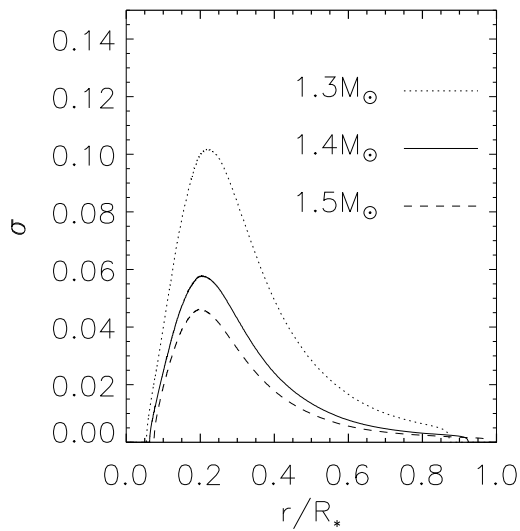,width=8cm}}
\caption{Profile of $\sigma(r)$ for three different stellar masses at age 
300Myr.
 The 1.3$M_{\odot}$, 1.4$M_{\odot}$ and 1.5$M_{\odot}$ stars are assumed to be 
rotating with the velocities $8 \times 10^{-5}$rad/s, $1.3 \times 10^{-4}$rad/s, 
and $1.5 \times 10^{-4}$rad/s respectively. 
The stellar models are computed using a standard stellar evolution code 
developed by Bodenheimer et al. (2007).}
\label{fig:sigma} 
\end{figure}

It is important to note that $\sigma$ depends on $\Omega_\star$, 
so that the effect of stratification (in the sense defined above) 
is strongly reduced for 
rapidly rotating stars. Figure \ref{fig:sigma} shows 
an estimate of $\sigma$ for various stars in the mass range 
$1.3M_\odot-1.5M_\odot$ 
at age 300Myr. In all cases, $\sigma$ 
remains well-below unity showing that large-scale mixing of the radiative zone
by gyroscopic pumping could be significant for these stars.

\begin{figure}[h]
\centerline{\epsfig{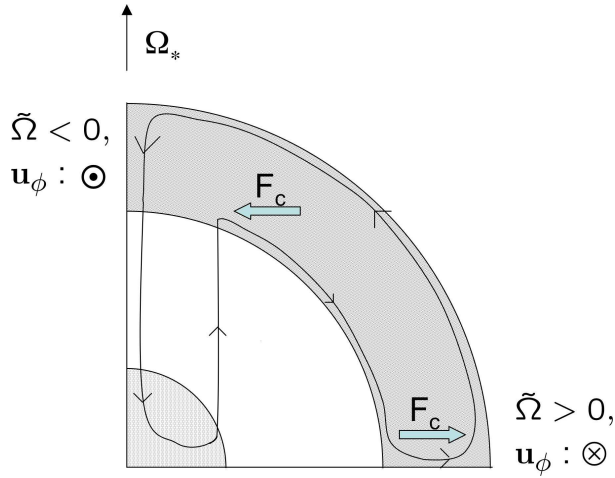}}
\caption{Illustration of the expected flow pattern for rapidly 
rotating stars with two convective zones. Gyroscopically pumped flows
from one convection zone can enter the radiative zone and return 
within the second convective zone where the Taylor-Proudman constraint
does not apply (see main text for detail). Note the role of the tangent cylinder
in such a model in delimiting mixed regions from non-mixed regions. }
\label{fig:gyro2} 
\end{figure}

Crucially, however, GAA09 showed that even in the very weakly stratified limit 
($\sigma \ll 1$), a second condition needs to be satisfied for the 
pumped flows to penetrate into the nearby radiative 
region. Indeed, this limit corresponds to the case where the system's 
dynamics are dominated by the balance
between the perturbation to the pressure gradient and the Coriolis force.
This well-known situation is called the ``Taylor-Proudman'' state, and 
(in the Boussinesq approximation) implies that all components of the 
velocity field are strongly 
constrained to be constant along the rotation axis. 
GAA09 showed that the Taylor-Proudman constraint 
 can prohibit flow generated within the outer convection zone from 
entering the underlying radiative zone. 
Indeed, by mass conservation, any flows entering the radiative zone  
must somehow return to the convection zone. But such return flow would
necessarily require breaking away from the Taylor-Proudman state. 
Hence, if there exists a region within the radiative zone 
where the Taylor-Proudman constraint is broken, then this region provides a 
channel through which the pumped flows can return, 
as illustrated in Figure \ref{fig:gyro2}. If such a region does not exist, 
the meridional flows instead return within the convection zone causing 
negligible mixing in the radiative zone.

Since the Taylor-Proudman constraint is broken whenever there exist
additional stresses of amplitude comparable with the Coriolis force, various
mechanisms can be invoked. Of particular interest are 
Maxwell stresses in the presence of 
small- or large-scale magnetic fields (see Gough \& McIntyre 1998, 
for example), and turbulent stresses within another
convective region. It is the latter we are mostly interested in studying in 
this work, namely the case of stars with a convective envelope {\it and} a
convective core.
 
This paper is organized as follows. We first present a fairly exhaustive
study of gyroscopic pumping for rapidly rotating stars 
(in the limit where $\sigma < 1$). In \S\ref{sec:model} we lay out
the general formulation of the problem and the assumptions made. Note that the 
basic model used in this paper is largely inspired from the work of 
GAA09 but extends it to the case of multiple convective regions and uses a 
more generally applicable formalism for the fluid dynamics, 
namely the anelastic approximation. 
In \S\ref{sec:cart} we first solve the problem analytically in 
a Cartesian coordinate system, using a much simplified stellar model. 
This exercise provides insight onto how the meridional flow velocities induced
by gyroscopic pumping scale with the forcing 
mechanism (in this case the differential rotation of the inner and/or outer 
convection zone), with the stratification within the radiative region, 
and with the system's geometry (i.e. the respective widths of the inner 
and outer convective regions).
In \S\ref{sec:spher}, we then apply the same model in a two-dimensional
spherical geometry, using a more realistic 
stellar model as the background state. By comparing 
Cartesian-model results with equivalent spherical-model results, 
we deduce a very simple rule to go from one to the other. This rule
is particularly useful since one-dimensional 
Cartesian model numerical solutions can be obtained in a tiny 
fraction of the time necessary to integrate the two-dimensional 
spherical case, and can also be pushed to true stellar parameter values 
(which cannot be done in two dimensions).

Next, we present a simple application of this theory to estimate 
the surface abundances of lithium (Li) and beryllium (Be)
in young Main Sequence stars as a function of their age and mass. 
More precisely, we are interested in young stars 
in the mass range of $1.3M_\odot-1.5M_\odot$, which have the well-known 
property of being significantly depleted in Li and Be in their surface layers  
as compared with slightly more and slightly less massive stars 
(Boesgaard \& Tripico 1986; see Boesgaard 2005 and Anthony-Twarog et al. 
2009 for reviews). These so-called ``Li-dip'' stars are unique in the sense 
that they have two significant convection zones (one inner and one outer)
which, as described earlier, could promote large-scale mixing between the 
surface and the interior by gyroscopic pumping. 
In \S\ref{sec:lidip}, we show that the Li and Be depletion 
rates as induced by gyroscopic pumping for Li-dip stars can be very significant. 
For reasonable model assumptions, the cool (low-mass) side of the dip 
is readily explained by gyroscopic pumping, while depletion fractions on the 
hot (high-mass) side are much larger than observed. When the diffusion of 
chemical species back into the outer convection zone 
(by overshooting motions for example)
is taken into account, good agreement between the model and the data is 
achieved on both sides of the dip. Finally, we conclude in 
\S\ref{sec:ccl} by summarizing our main results and discussing future prospects. 

\section{The model} 
\label{sec:model}

In this section we briefly derive the model equations used throughout this 
paper. We consider a star of radius $R_\star$, rotating with a mean 
angular velocity $\Omega_\star$. In all that follows, we assume that 
the system is in a quasi-steady state, non-magnetic, and axially symmetric. 
While these assumptions are probably over-simplistic, they help us 
narrow down gyroscopic pumping to its essence. 

The star considered can have up to two convective regions: 
a convective core, extending from the center of the star to a first
radiative--convective interface located at the radius 
$r =r_{\rm in}$; and a convective envelope, 
located between the outer radiative--convective interface at 
$r =r_{\rm out}$ and the surface. Note that the stellar surface and 
both interfaces are assumed to be perfectly spherical. 

GAA09 studied large-scale meridional flows within the solar interior 
using the Boussinesq 
approximation. This approximation treats the thermodynamical quantities 
as the sum of a weakly varying background plus small perturbations. One 
may argue in favor of its use in the solar radiative zone, which does
not span too many density scaleheights. Here, however, we aim to model 
stars with very thin outer convective regions, 
in which case the radiative zone extends nearly all the way to 
the stellar photosphere. We must therefore switch to using the more general 
anelastic approximation instead, which allows for a more strongly varying 
background stratification. 

The anelastic approximation implicitly
assumes that all velocities are small compared with the local sound 
speed, and that thermodynamical perturbations are small compared
with the equivalent background quantities. We thus define $\bar p(r)$, 
$\bar \rho(r)$, $\bar T(r)$ and $\bar s(r)$ as the spherically symmetric 
background pressure, density, temperature and entropy profiles, 
and the equivalent $\tilde{p}$, $\tilde \rho$, $\tilde T$ and $\tilde{s}$ 
as two-dimensional perturbations to these quantities.  
If the equation of state is assumed to be that of a perfect gas 
(which is an acceptable approximation for the purpose of this work), 
then the perturbations are related by 
\begin{equation}
\frac{\tilde{p}}{\bar p} = \frac{\tilde{\rho}}{\bar \rho} + \frac{\tilde{T}}{\bar T}\mbox{   ,  }
\end{equation}
neglecting for simplicity the dependence on the chemical gradients. 

The momentum, thermal energy and mass 
conservation equations describing the dynamics of the interior flows and
thermodynamical perturbations, in the anelastic approximation, are:
\begin{eqnarray}
\label{eq:initeqs}
\bu \cdot \nabla \bu + 2 {\bf \Omega}_\star \times \bu = - \frac{\nabla \tilde p}{\bar \rho} + \frac{\tilde{\rho}}{\bar \rho} \bar \bg  +  \frac{1}{\bar \rho} \nabla \cdot \Pi \mbox{   ,  } \nonumber \\
\bar\rho \bar c_p \bar T \bu \cdot \nabla \bar s  = \nabla \cdot ( \bar k_T \nabla \tilde{T}) - \nabla \cdot F_T \mbox{   ,  }\nonumber \\
\div(\bar \rho \bu) = 0 \mbox{   ,  }
\end{eqnarray}
where $\bu$ is the velocity field expressed in a frame rotating with 
angular velocity $\Omega_\star$, $\bar \bg = (0,0,-\bar g)$ is gravity, 
$\Pi$ is the viscous stress tensor, 
$\bar c_{\rm p}$ is the specific heat at constant pressure, 
$\bar k_T = \bar \rho \bar c_{\rm p} \bar \kappa$ is the 
thermal conductivity and finally $F_T$ is the 
turbulent heat flux (in the convective regions). Note that we have neglected 
any distortion of the star caused by the centrifugal force, as well 
as perturbations to the gravitational field. These assumptions
suppress the well-known {\it global} Eddington-Sweet flows
(see Spiegel \& Zahn 1992). 

Following GAA09, we model the inertial term of the momentum equation
in the convective regions by the sum of a turbulent viscosity plus a 
linear drag term driving the system toward a differentially rotating profile: 
$-( \bu -u_{\rm cz} \ephi )/\tau$, where 
$u_{\rm cz} \ephi $ is the assumed/observed azimuthal velocity profile 
in the convective zones and where $\tau$ is the local 
convective turnover timescale (which varies with depth). 
This drag term is introduced to ``mimic''
the effect of turbulent convection on driving differential rotation. It
is only significant in the convection zones, and drops to zero in 
the radiative region. The very slow flow velocities expected in 
the radiative zone justify neglecting the various nonlinear terms in the 
momentum and heat equation there. 
We replace the turbulent heat advection term by a 
turbulent diffusivity, which is assumed to be 
very large in the convection zones and rapidly tends to zero otherwise.

The resulting model equations, which we use throughout this work unless 
otherwise specified, are therefore:
\begin{eqnarray}
\label{eq:gov_eqs}
2 {\bf \Omega}_\star \times \bu = - \frac{\nabla \tilde p}{\bar \rho} +\frac{\tilde{\rho}}{\bar \rho}\bar \bg  +  \frac{1}{\bar \rho} \nabla \cdot \left(\Pi + \Pi_{\rm turb} \right)  - \frac{\bu - u_{\rm cz}\ephi}{\tau} \mbox{   ,  } \nonumber \\
\frac{\bar\rho \bar c_p \bar T \bar N^2}{g} \bu \cdot \be_r   = \nabla \cdot ( (\bar k_T+k_{\rm turb})  \nabla \tilde{T})\mbox{   ,  } \nonumber \\
\frac{\tilde{p}}{\bar p} = \frac{\tilde{\rho}}{\bar \rho} + \frac{\tilde{T}}{\bar T} \mbox{   ,  }\nonumber \\
\div(\bar \rho \bu) = 0 \mbox{   ,  }
\end{eqnarray}
where $\Pi_{\rm turb}$ is similar to the microscopic stress tensor, but
 using a turbulent viscosity instead. Note that we have rewritten the 
background heat advection term to emphasize the dependence on the 
buoyancy frequency $\bar N$ (see Spiegel \& Zahn 1992, for example). 

\section{A simplified Cartesian model}
\label{sec:cart}

Much can be learned about the dynamics of stellar interiors by first 
studying a simplified problem in Cartesian geometry (the ``planar star'' 
approximation, see Garaud \& Brummell 2008 and GAA09 for example). 
Equations in this geometry can usually
be solved analytically to gain insight 
into the physical processes at play. 
They often reveal important scaling laws governing the solutions, and finally, 
are rarely more than an ``order one''
geometrical factor away from more realistic solutions in spherical 
geometry (in fact we prove this in \S\ref{sec:spher}).
Our primary goals in this section are therefore 
not quantitative. Rather, we aim to 
determine, qualitatively, how deeply the meridional 
flow velocities generated by gyroscopic pumping 
penetrate into the radiative zone, and characterize how 
their amplitude scales with the system parameters. 

Since the Cartesian model 
solutions obtained are knowingly off by a factor of order unity anyway, 
we further simplify the equations, in this section only, with the following 
substitutions:
\begin{eqnarray}
\frac{1}{\bar \rho} \div (\Pi + \Pi_{\rm turb}) \rightarrow (\bar \nu+\nu_{\rm turb}) \grad^2 \bu \mbox{   ,  }  \nonumber \\
\nabla \cdot \left( (\bar k_T + k_{\rm turn}) \nabla \tilde{T} \right) \rightarrow (\bar k_T + k_{\rm turb}) \nabla^2 \tilde{T}\mbox{   .  } 
\end{eqnarray}
It can be shown (through numerical integrations) 
that neither of these substitutions affect the scalings 
of the solutions in the case where both convection zones are present\footnote{The 
substitution of the stress tensor, however, modifies the nature of the viscous
boundary layers (the well-known Ekman layers, see for example Kundu 1990), 
which are relevant in the case of stars with a single 
convection zone only.}. 
Finally, and following Spiegel \& Zahn (1992) we 
neglect in this section, for analytical simplicity, the 
pressure perturbations in the linearized equation of state so that: 
\begin{equation}
\frac{\tilde{\rho}}{\bar \rho} = - \frac{\tilde{T}}{\bar T} \mbox{  . }
\end{equation}

\subsection{Model setup and non-dimensional equations}
\label{sec:modeleq}

\begin{figure}[h]
\centerline{\epsfig{file=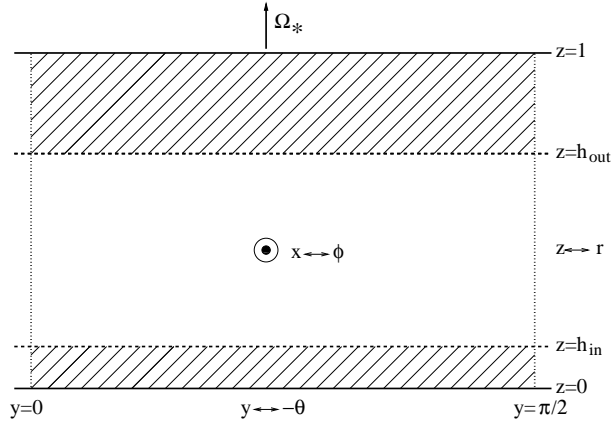,width=8cm}}
\caption{Cartesian model geometry and intended correspondence with the 
spherical case. The shaded area marks the convective regions, where forcing 
is applied. The $y=0$ and $y=\pi/2$ lines mark the ``poles'' 
and the ``equator''. The system is assumed to be periodic with period 
$\pi$ in the $y-$direction. 
}
\label{fig:model} 
\end{figure}

As in GAA09, we consider a Cartesian coordinate system $(x,y,z)$ with $z$ 
aligned with both gravity and with the rotation axis. 
The $x$-direction represents
the azimuthal direction, while the $y$-direction is equivalent to minus 
the co-latitude. Distances are normalized to the stellar radius $R_\star$, 
so that the stellar interior is in the interval $z \in [0,1]$,
the inner convection 
zone spans the interval $[0,z_{\rm in}]$ and the outer convection 
zone spans $[z_{\rm out},1]$. Figure \ref{fig:model} illustrates 
the geometry of the Cartesian system.
We assume that the star is axially symmetric, i.e. 
independent of $x$, and periodic in $y$ on the interval 
$[0,\pi]$ (representing the two ``poles'') with equatorial symmetry 
(i.e. symmetric about $y = \pi/2$).

The background stellar model is chosen to be very simple, so that 
analytical solutions of the problem can easily be found. We take
\begin{itemize}
\item $\bar \rho(z) = \rho_c e^{-z/D_\rho}$ and $\bar T(z) = T_c e^{-z/D_T}$, 
\item $\bar \nu$, $\bar \kappa$ and $\bar g$ are constant.
\end{itemize}
Note that fits to actual stellar models show that the non-dimensional density 
and temperature scaleheights typically 
satisfy the inequalities $D_\rho \ll D_T < 1$. 

The global velocity field in the rotating frame
is ${\bf u } = (u,v,w)$ and flow velocities are normalized to  
$R_\star \Omega_\star$. In this framework, the unit timescale 
is $\Omega_\star^{-1}$. 
Density and temperature perturbations are normalized by $\rho_c$ and 
$\epsilon T_c$ respectively, where $\epsilon = R_\star \Omega_\star^2 /\bar g$ 
is the ratio of the centrifugal force to gravity. Pressure perturbations are 
normalized to $\rho_c R^2_\star \Omega_\star^2$. 
The set of equations (\ref{eq:gov_eqs}) 
then simplifies to the non-dimensional system:
\begin{eqnarray}
\label{eq:mod_eqs}
2 \ez \times \bu = - \nabla \tilde{p} \, e^{z/D_\rho} +  \tilde{T} e^{z/D_T} \ez + (E_{\nu} + E_{\nu, {\rm turb}}) \grad^2 \bu - \frac{\bu - \bu_{\rm cz}}{\tau} \mbox{   ,  } \nonumber \\
\frac{\bar N^2}{\Omega_\star^2} e^{-z/D_T} w = (E_\kappa + E_{\kappa,{\rm turb}} ) \nabla^2 \tilde{T} \mbox{   ,  } \nonumber \\
\div( e^{-z/D_\rho}\bu) = 0 \mbox{   ,  } 
\end{eqnarray}
where we have introduced a series of standard parameters, namely 
the Ekman number
\begin{equation}
\Enu = \frac{\bar \nu}{R_\star^2 \Omega_\star} \mbox{  ,  } 
\end{equation}
and the equivalently defined $E_{\nu,{\rm turb}}$, as well as 
\begin{equation}
E_\kappa = \frac{\bar \kappa }{ R_\star^2 \Omega_\star } \mbox{  ,  } 
\end{equation}
which is actually an inverse Peclet number, and the equivalently defined 
$E_{\kappa,{\rm turb}}$. Note that the microscopic diffusivities
normally vary with depth within a star, but are assumed
here to be constant for simplicity. Deep within the 
interiors of Hyades-age stars in the Li-dip mass range, 
\begin{equation}
 E_\nu \sim 10^{-17} - 10^{-15} \mbox{  ,  } 
 E_\kappa \sim 10^{-12} - 10^{-7} \mbox{  .  } 
\end{equation}

In order to fully specify the model, the functions $\bar N(z)$, 
$\tau(z)$, $u_{\rm cz}(y,z)$ and the turbulent diffusivity profiles 
must be selected. The buoyancy frequency is taken to be 
\begin{equation}
\bar N(z) = \frac{N_{\rm rz}}{2} \left[ \tanh\left( \frac{z_{\rm out}-z}{\Delta_{\rm out}} \right) +  \tanh\left( \frac{z-z_{\rm in}}{\Delta_{\rm in}} \right)  \right]\mbox{  ,  }
\end{equation}
so that $\bar N(z) \rightarrow 0$ in both convection zones, 
and $\bar N(z) \simeq N_{\rm rz}$ in the radiative zone. 
The lengthscales $\Delta_{\rm in}$ and $\Delta_{\rm out}$ may be 
thought of as the respective thicknesses of the ``overshoot'' regions 
located near each of the two convection zones (and in this section are
taken to be equal to one another, for simplicity). 
For $\tau(z)$, we take
\begin{equation}
\label{eq:tau}
\tau(z)^{-1} = \frac{\Lambda_{\rm in}}{2} \left[1 + \tanh\left(\frac{z_{\rm in}-z}{\Delta_{\rm in}}\right)\right] + \frac{\Lambda_{\rm out}}{2} \left[1 +  \tanh\left(\frac{z-z_{\rm out}}{\Delta_{\rm out}}\right) \right] \mbox{   ,  } 
\end{equation}
where $\Lambda_{\rm in}$ and $\Lambda_{\rm out}$ are assumed to 
be constant and equal to the 
inverse of the (non-dimensional) convective turnover time in the 
relevant convection zone. 

Both convective zones may be differentially rotating. For mathematical 
simplicity again, we assume in this Cartesian model that their latitudinal 
dependence is similar. We therefore select 
the following functional form for $u_{\rm cz}(y,z)$:
\begin{equation}
u_{\rm cz}(y,z) = \hat u_{\rm cz}(z) e^{iky} \mbox{   ,  }
\label{eq:ucz}
\end{equation}
where $k = 2$ to guarantee equatorial symmetry. 
The radial profile $\hat u_{\rm cz}(z)$ is then chosen to be 
\begin{equation}
\hat u_{\rm cz}(z) =  \frac{\hat u_{\rm cz}^{\rm out}(z)}{2} \left[1+\tanh\left(\frac{z-z_{\rm out}}{\Delta_{\rm out}}\right)\right] + \frac{\hat u_{\rm cz}^{\rm in}(z)}{2} \left[1 + \tanh\left(\frac{z_{\rm in}-z}{\Delta_{\rm in}}\right)\right] \mbox{   .  }
\label{eq:uczprof}
\end{equation}
The functions $\hat u_{\rm cz}^{\rm out}(z)$ and $\hat u_{\rm cz}^{\rm in}(z)$ 
can a priori depend on depth (see GAA09 for example). 
By analogy, the non-dimensional turbulent diffusivity profiles are constructed as 
\begin{equation}
E_{\nu,{\rm turb}}(z) =  \frac{E^{\rm out}_{\nu,{\rm turb}}(z)}{2} \left[1+\tanh\left(\frac{z-z_{\rm out}}{\Delta_{\rm out}}\right)\right] + \frac{E^{\rm in}_{\nu,{\rm turb}}(z)}{2} \left[1 + \tanh\left(\frac{z_{\rm in}-z}{\Delta_{\rm in}}\right)\right]\mbox{   ,  }
\label{eq:eturbs}
\end{equation}
and similarly for $E_{\kappa,{\rm turb}}(z)$.

Projecting the model equations into Cartesian coordinates, using 
invariance in the $x-$direction and seeking periodic solutions in 
the form of $q(y,z) = \hat q(z) e^{iky}$ for each of the dependent 
variables yields:
\begin{eqnarray}
&&-2 \hat v =  (E_\nu + E_{\nu,{\rm turb}} )\left(\hat u_{zz} - k^2 \hat u \right) - \frac{\hat u-\hat u_{\rm cz}}{\tau} \mbox{   ,   }  \nonumber \\
&&2 \hat u = - ik \hat p e^{z/D_\rho}    + (E_\nu + E_{\nu,{\rm turb}} ) \left( \hat v_{zz} - k^2 \hat v \right) - \frac{\hat v}{\tau} \mbox{   ,   }  \nonumber \\
&&0 = - \hat p_z  e^{z/D_\rho}+  \hat T e^{z/D_T} +(E_\nu + E_{\nu,{\rm turb}} ) \left(\hat w_{zz} - k^2 \hat w \right) - \frac{\hat w}{\tau}  \mbox{   ,   }  \nonumber \\
&& \frac{\bar N^2}{\Omega_\star^2} e^{-z/D_T}   \hat w =(E_\kappa + E_{\kappa,{\rm turb}} )\left( \hat T_{zz}- k^2 \hat T \right) \nonumber \\
&& ik  \hat v e^{-z/D_\rho}  + ( e^{-z/D_\rho}\hat w)_z = 0 \mbox{   .   }
\label{eq:maineqs}
\end{eqnarray}
where the subscript $z$ denotes a derivative with respect to $z$.

Finally, we need to specify an adequate set of boundary conditions for 
the system. The two boundaries at $z=0$ and $z=1$ are assumed to be 
impermeable ($\hat w = 0$), stress-free ($\hat u_z = \hat v_z =0$), 
and the temperature perturbations are assumed to be zero. Note that 
as long as the system boundaries are located in a convective region,
the actual choice of boundary conditions has little influence on the 
result.

\subsection{Solution and interpretation of the model}
\label{sec:cartsol}

The set of equations (\ref{eq:maineqs}) and associated boundary conditions
(see above) can be solved analytically when 
the overshoot regions are very thin compared with the depths of the respective
convective zones. The complete 
derivation of the solution is fairly straightforward although algebraically
cumbersome. It is detailed in Appendix A: 
exact solutions are derived in each of the three regions
$[0,z_{\rm in}]$, $[z_{\rm in},z_{\rm out}]$ and $[z_{\rm out},1]$, 
and matched to one another across the radiative--convective 
interfaces at $z_{\rm in}$ and $z_{\rm out}$ respectively. In what follows,
we discuss the most important outcome of this analysis, namely the 
prediction of the gyroscopically pumped mass flux mixing the radiative zone. 

\subsubsection{General behavior of the model solutions.}
\label{sec:analytical}

As found by GAA09, in the limit where $\sigma < 1$ 
the meridional flows generated in the convective regions can penetrate 
deeply into the nearby radiative zone. A very simple way of seeing this is 
to note that in this limit, the $x-$component of the momentum equation in 
(\ref{eq:maineqs}) reduces to $\hat v = O(E_\nu)$ in the radiative zone 
which then implies, by mass conservation, that
\begin{equation}
(e^{-z/D_\rho}\hat w)_z = O(E_\nu) \mbox{  .  } 
\end{equation}
Hence the non-dimensional vertical mass flux mixing the radiative zone, 
$\bar \rho \hat w$, is constant along the rotation axis: 
\begin{equation}
\bar \rho \hat w = W_{\rm rz}  \mbox{  ,  } 
\end{equation}
and spans the entire region, extending from one convection zone to 
the other as drawn in Figure \ref{fig:gyro2} for example. 

The details of the calculation of the pumped mass flux $W_{\rm rz}$, even for 
this simplified stellar model, are fairly complicated and are presented 
in Appendix A. 
In the limit where the depths of both convection zones $d_{\rm in} = z_{\rm in}$ and $d_{\rm out} = 1-z_{\rm out}$ 
are small compared with the stellar radius 
(which is true for most stars in the Li dip), we show that 
\begin{equation}
W_{\rm rz} = \frac{2}{k} \frac{P}{\frac{N_{\rm rz}^2}{\Omega_\star^2}\frac{G}{E_\kappa}   - \frac{1}{k^2} \left( \frac{4 + \Lambda_{\rm in}^2 }{\Lambda_{\rm in} d_{\rm in}} + \frac{4 + \Lambda_{\rm out}^2 }{\Lambda_{\rm out}d_{\rm out}} \right) }\mbox{  ,  } 
\label{eq:wrzconst}
\end{equation}
where $P$ is the ``pumping term'' 
\begin{eqnarray}
P &=& \bar \rho(z_{\rm out}) u_{\rm cz}^{\rm out}(z_{\rm out}) - \bar \rho(z_{\rm in}) u_{\rm cz}^{\rm in}(z_{\rm in})  +  d_{\rm out} \left. \frac{\dd (\bar\rho u^{\rm out}_{\rm cz})}{\dd z}\right|_{z=z_{\rm out}}  + d_{\rm in}\left. \frac{\dd (\bar\rho u^{\rm in}_{\rm cz})}{\dd z}\right|_{z=z_{\rm in}} \mbox{  ,  }  
\end{eqnarray}
and where $G$ is the following fairly obscure factor:
\begin{eqnarray}
&& G = L^2(z_{\rm out} - z_{\rm in}) \nonumber \\
&-& k L^4  \left[1- \frac{2L}{d_{\rm out}}  +  \frac{2L}{ d_{\rm out}} e^{-d_{\rm out}/L} \right]  \left[ \frac{e^k}{\sinh(k)}  \left( 1+ \frac{d_{\rm out}}{L} \right) - \frac{e^{-(z_{\rm out}-z_{\rm in})/L}}{\sinh(k)} \left(1 - \frac{d_{\rm in}}{L} \right)   - \frac{1}{kL} \right]  \nonumber \\
&+& kL^4 \left[  1 + \frac{2L}{d_{\rm in}} - \frac{2L}{z_{\rm in}}  e^{z_{\rm in}/L} \right]  \left[  \frac{e^{(z_{\rm out}-z_{\rm in})/L}}{ \sinh(k)} \left( 1+ \frac{d_{\rm out}}{L} \right) - \frac{e^{-k}}{\sinh(k)} \left(1 - \frac{d_{\rm in}}{L} \right)  -   \frac{1}{kL} \right] \mbox{ ,}
\end{eqnarray}
where 
\begin{equation}
L^{-1} = D_\rho^{-1} - D_T^{-1} \mbox{ .}
\end{equation}
Note that $G$ is found to be negative for all reasonable parameter values, so that the denominator
of (\ref{eq:wrzconst}) never vanishes. Our analytical solution is 
easily verified by comparison with numerical solutions of the full 
set of equations (\ref{eq:maineqs}), as shown in \S\ref{sec:comparecart}.
However, let us first attempt to understand the meaning of (\ref{eq:wrzconst})
on physical grounds.

\subsubsection{Interpretation of the dependence of the pumped mass flux on physical parameters}
\label{sec:physmeaning}

This expression for $W_{\rm rz}$ calculated in \S\ref{sec:analytical}
can be interpreted more easily in two different asymptotic limits. 

\paragraph{The unstratified limit.} When $N_{\rm rz} \rightarrow 0$ 
(the unstratified limit), $W_{\rm rz}$ takes the simpler 
form\footnote{Note that this 
expression for $W_{\rm rz}$ can be derived directly, and much more easily, 
by considering an unstratified system in the first place 
(ignoring the buoyancy term in the momentum equation, 
taking $\bar \rho$ and $\bar T$ constant, and ignoring the thermal 
energy equation).}:  
\begin{equation}
|W_{\rm rz}| =  \frac{2}{k} \frac{P}{\frac{4 + \Lambda_{\rm in}^2 }{\Lambda_{\rm in} d_{\rm in}} + \frac{4 + \Lambda_{\rm out}^2 }{\Lambda_{\rm out}d_{\rm out}}} \mbox{  . }
\end{equation}
We see that $W_{\rm rz}$ is roughly of the 
order of the pumping term, times a factor which depends only on the 
respective properties (depth, convective turnover time) of the convective zones. 
Our main conclusion is that the mass flux into the radiative zone 
appears to go to zero\footnote{In practice, if one of the 
convection zones vanishes entirely ($d_{\rm in} = 0$ or $d_{\rm out} = 0$) 
then the analysis presented in Appendix A 
is no longer valid. It can be shown instead (analytically and numerically) 
that the meridional flow amplitudes do not 
entirely drop to zero but instead drop to the level of Ekman (viscous) 
flows and depend sensitively on the boundary conditions. Meanwhile,
if $\Lambda_{\rm in} = 0$ {\it or} $\Lambda_{\rm out} = 0$, but the turbulent 
stresses $E_{\nu,{\rm turb}}^{\rm in}$ and $E_{\nu,{\rm turb}}^{\rm out}$ 
are non-zero, then mixing of the radiative zone by the pumped flows can 
still be effective. Indeed, gyroscopic pumping from one of the two 
convective zones is still effective, and the second provides the 
return pathway for the flows. This effect is 
not expressed in (\ref{eq:wrzconst}), since our analytical derivation  
ignores for simplicity the effect of the turbulent
diffusion term compared with the relaxation term.} if one 
of the convection zones vanishes, either by becoming
vanishingly thin ($d_{\rm in}$ or $d_{\rm out} \rightarrow 0$), 
or if the associated convective stresses become negligible 
($\Lambda_{\rm in}$ or $\Lambda_{\rm out}  \rightarrow 0$).

This important result can in fact be easily understood
in the light of the work of GAA09 described in \S\ref{sec:intromix}. 
Indeed, in the unstratified case, radiative zone flows must satisfy 
the Taylor-Proudman constraint. Any flows generated by gyroscopic pumping in one
convection zone can only enter the radiative zone if there is a return path  
at the other end. If the second convective zone vanishes, this return path is no longer 
available. The flows instead return within the existing convection zone, and do not
mix the radiative region significantly. The stratified case with $\sigma < 1$ is very similar. 

Based on these very simple considerations, we can therefore expect that the 
overall mixing rate in the star resulting
from gyroscopic pumping must reach a maximum for a given stellar mass between 
$1 M_\odot$ (no or negligible convective core) and $1.8 M_\odot$ 
(no or negligible convective envelope). This simple idea motivated our 
study of the Li dip (see \S\ref{sec:lidip}), although, as we shall show,
the real problem is much more subtle. 

\paragraph{The stratified case.} 
The unstratified limit discussed above is of course artificial. In real
stars $N_{\rm rz} \neq 0$ and one must instead compare the 
two terms in the denominator of (\ref{eq:wrzconst}) to one another. 
In the limit where the first term is much larger than the second then
\begin{equation}
|W_{\rm rz}| \simeq  \frac{2P}{k} \frac{E_\kappa}{G} \frac{\Omega_\star^2}{N_{\rm rz}^2}  \mbox{   . }
\label{eq:esflows}
\end{equation}
The flow velocities pumped into the
radiative zone are now found to follow a local Eddington-Sweet scaling law, 
which is not surprising since 
we are looking at quasi-steady flows in a stratified fluid driven by rotational 
forcing. Such solutions were already found by Spiegel \& Zahn (1992) for example in the case
of the Sun. Naturally, local Eddington-Sweet
flows are much slower than the flows pumped through each individual 
convective zone, although they could still provide significant sources of 
mixing in fairly rapidly rotating stars. 

The effect discussed in the unstratified case, namely 
the complete suppression of $W_{\rm rz}$ in the limit where 
one of the convection zone disappears, still occurs 
but only when 
\begin{equation}
\frac{\sigma^2 G}{E_\nu} = \frac{N_{\rm rz}^2}{\Omega_\star^2}\frac{G}{E_\kappa}  \ll \frac{1}{k^2} \left( \frac{4 + \Lambda_{\rm in}^2 }{\Lambda_{\rm in} d_{\rm in}} + \frac{4 + \Lambda_{\rm out}^2 }{\Lambda_{\rm out}d_{\rm out}} \right) \mbox{   .}
\end{equation}
Since $N_{\rm rz}^2/E_\kappa \Omega_\star^2$ is typically very large because $E_{\kappa}$ is very small, this limit is only relevant for {\it extremely} 
thin or weak convective regions. 

Based on these considerations, we can now re-interpret (\ref{eq:wrzconst}) 
in the following way: $W_{\rm rz}$ has two strict upper limits: 
a first upper limit, which arises from mechanical constraints (by the 
driving force and the Taylor-Proudman constraint), and a second upper limit
which arises from the thermal stratification of the system (which can 
strongly suppress radial flows). The actual value of $W_{\rm rz}$ mixing the radiative zone
is the smaller of the two, thus defining two different regimes, the ``unstratified'' regime
(where $W_{\rm rz}$ is mechanically constrained) and the ``weakly stratified'' regime
(where $W_{\rm rz}$ is thermally constrained).

\subsubsection{Comparison with full solutions of the governing equations}
\label{sec:comparecart}

In this section we verify the analytical solution for $W_{\rm rz}$ 
expressed in (\ref{eq:wrzconst}) by comparison 
with numerical solutions of the same equations. 
Numerical solutions are obtained 
by integrating the two-point boundary value problem (\ref{eq:maineqs}) 
with associated boundary conditions, background state
and forcing as specified in \S\ref{sec:analytical}. 

\begin{figure}[h!]
\centerline{\epsfig{file=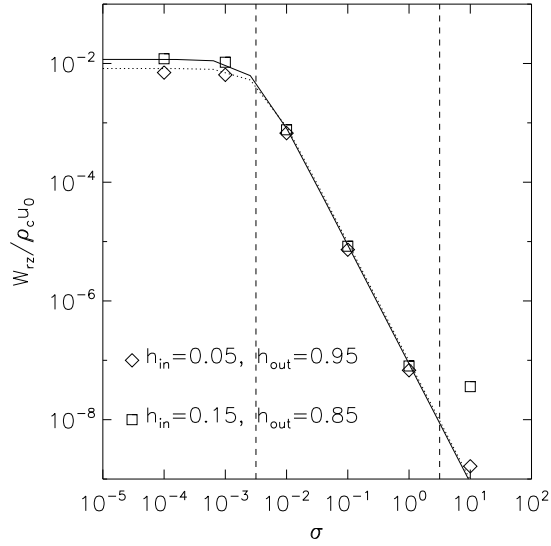,width=8cm}}
\caption{Comparison between the numerical and analytical results for 
$W_{\rm rz} = \bar \rho \hat w$. For each selected value of $\sigma$ 
ranging from $10^{-4}$ to 10, two sets of simulations are shown with 
different convection zone depths: the case of thinnest 
convection zones ($z_{\rm in}=0.05$, $z_{\rm out} = 0.95$) is shown as 
diamond symbols, while ($z_{\rm in}=0.15$, $z_{\rm out} = 0.85$) 
is shown as squares. 
The corresponding analytical solutions are shown as dotted and solid lines
respectively. 
For $\sigma < 1$ the numerical and analytical results fit very well for 
all cases, showing that the approximate formula for $W_{\rm rz}$ 
(see equation (\ref{eq:wrzconst})) applies. For $\sigma > 1$ 
the formula doesn't do so well, as expected.}
\label{fig:wrz}
\end{figure}

A direct comparison of the analytical and numerical solutions for 
$W_{\rm rz}$ is shown in Figure \ref{fig:wrz}, 
for a wide range of simulations. For all the calculations presented, 
the microscopic diffusion parameters $E_\nu = 10^{-7}$ and the Prandtl number 
Pr = $E_\nu/E_\kappa = 10^{-2}$ are fixed. The turbulent diffusivities
are set to 0 for simplicity. For the forcing by the differential
rotation, as expressed in (\ref{eq:uczprof}), we take 
$\hat u_{\rm cz}^{\rm out}(z) = \hat u_{\rm cz}^{\rm in}(z) = u_0$  and 
$\Lambda_{\rm in} = \Lambda_{\rm out} = 10$. Note that the problem is linear so the value of $u_0$ is irrelevant. The overshoot depths are taken 
to be $\Delta_{\rm in} = \Delta_{\rm out} = 10^{-4}$.
The background density and temperature
scaleheights are selected to be $D_\rho = 0.075$ and $D_T = 0.275$, so that 
$L = 0.1$. Note that these scaleheights are a good approximation to the 
true density and temperature scaleheights in the radiative zones of stars 
in the $1.3M_\sun - 1.5M_\sun$ range.  The non-dimensional buoyancy frequency 
of the radiative zone $N_{\rm rz}/\Omega_\star$ is varied for each set of simulations, leading to 
$\sigma$ ranging from $10^{-4}$ to 10. Finally, each symbol corresponds 
to a particular geometry of the system, with different possible pairs of 
values $(z_{\rm in},z_{\rm out})$ as shown.  

For each simulation the quantity $\bar \rho \hat w$ is measured 
from the numerical solution at $z= 0.5$. Note that
for $\sigma<1$,  $\bar \rho  \hat  w$ is found to be constant across much of the
radiative zone (as expected), so exactly where $\bar \rho  \hat  w$ is measured does not matter
much. For $\sigma \ge 1$, $\bar \rho  \hat  w$ is no 
longer constant, so the specific height $z= 0.5$ is selected for consistency
across simulations.
We find that the analytical solution (\ref{eq:wrzconst}), shown in the 
dotted and solid lines, fits the numerical results very well as long as $\sigma < 1$. 
The mismatch for $\sigma> 1$ was expected because we can no longer assume that 
$\bar \rho \hat w$ is constant across the radiative zone 
(see GAA09 for example), while 
this is key to the derivation of (\ref{eq:wrzconst}).
Figure \ref{fig:wrz} also shows the two regimes discussed in \S\ref{sec:comparecart}. 
In the ``unstratified limit'', for $\sigma \rightarrow 0$, we observe a plateau in $W_{\rm rz}$
which sets the largest achievable value for flows in the 
interior. As the degree of stratification increases, the flows driven into the 
radiative zone are slowed down by the local thermal stratification, a
phenomenon which becomes more pronounced as $N_{\rm rz}$ increases 
(equivalently, as $\sigma$ increases). The flow velocities 
in this ``weakly stratified'' regime scale as $\sigma^{-2}$, which
is equivalent to the aforementioned local Eddington-Sweet 
scaling since $E_\nu$ and Pr are constant. Note that in this regime, the flow pattern 
still spans the entire region with constant $\bar \rho \hat w$. 




\subsection{Summary}

To summarize this section, we have used a simple toy model to 
study how gyroscopic pumping by convective zone stresses induces 
large-scale meridional fluid motions. We found
that the pumping indeed drives significant large-scale flows within 
the convection zone(s) (see GAA09). 
Furthermore, in some circumstances, a fraction of the 
pumped mass flux may enter the adjacent radiative zone, and 
induce a circulation of material with the following properties:
\begin{itemize}
\item Convection zone flows penetrating into the nearby radiative zone are
exponentially damped on a lengthscale $R_\star/\sigma$ where $\sigma$ is given 
by equation (\ref{eq:sigmadef}). If $\sigma < 1$ (which is the case for
most rapid rotators) the flows can {\it potentially} mix the entire star 
(Garaud \& Brummell 2008, GAA09). 
\item In the limit $\sigma < 1$ the pumped mass flux 
{\it within the radiative zone} $\bar \rho \hat w$ 
is mechanically constrained to be constant along the rotation axis. 
\item The fraction of the gyroscopically pumped mass flux which does enter the 
radiative zone (by contrast with the mass flux which returns immediately 
within the driving convection zone) is capped by the lowest of two 
constraints: a 
mechanical constraint, which crucially depends on the presence of another 
source of stresses somewhere else within the system to enable flows to 
return to their point of origin (see previous sections for detail, 
as well as GAA09), and 
a thermal constraint, which limits the flow velocities to local 
Eddington-Sweet velocities. Note that this second constraint was 
not discussed by GAA09, but turns out to be the most relevant 
one for most stars.
\end{itemize}

The efficiency of gyroscopic pumping on mixing 
stellar interiors thus depends on many factors, including the background 
stellar structure, the thermal diffusivity, the stellar rotation rate and the 
nature of the convective zone stresses. However, thanks to the analytical 
formula (\ref{eq:wrzconst}), we now have a reasonably clear picture of precisely 
how all of these factors influence the amplitude of the gyroscopically 
pumped mass flux within a star.

\section{From Cartesian to Spherical models}
\label{sec:spher}

The next step of this investigation is to move to more realistic numerical simulations 
of the problem in a spherical geometry. We have two goals in this endeavour. The first
is to understand the effect of the spherical geometry on the solutions. 
In particular, we are interested in the dichotomy
between the regions located respectively within and outside of the cylinder tangent 
to the convective core and aligned with the rotation axis (see Figure \ref{fig:gyro2}). 
The second goal is much more quantitative, and is to extract (if possible) simple laws relating 
the calculated mass flux $W_{\rm rz}$ in the Cartesian case to its equivalent in the spherical case. 
Since Cartesian solutions are much easier to calculate than spherical geometry solutions, 
a simple rule to go from one set of solutions to the other could prove
particularly useful later. 

\subsection{Model description}
\label{sec:sphermod}

The spherical model used is very similar to the model presented in GAA09.
The salient points are repeated here for completeness. 

We consider a spherical coordinate system $(r,\theta,\phi)$ where 
$\theta = 0$ denotes the rotation axis and $\theta = \pi/2$ marks the equator. 
The governing equations 
are the original ones (\ref{eq:gov_eqs}) derived in \S\ref{sec:model}. 
Since we are principally interested in studying the effects
of the spherical geometry on the model predictions for the large-scale 
flow amplitudes, we consider ``hypothetical stars'' instead of 
real stellar models. This largely facilitates the comparison between 
the various simulation outputs, and enables us to focus on how the 
model depends on specific control parameters.

In all cases, the ``star'' used is a solar-type star 
($R_\star = R_\sun$, $M_\star = M_\sun$), and the background thermodynamical 
quantities such as density, pressure, and temperature 
($\bar{\rho}(r)$, $\bar{p}(r)$ and 
$\bar{T}(r)$ respectively) are extracted from Model S 
of Christensen-Dalsgaard et al. (1996). However, 
to model cases with various size convection zones, we create {\it artificial} 
profiles of the buoyancy frequency by using the expression
\begin{eqnarray}
\bar N^2(r) = N^2_{\rm rz} \sin \left( \frac{ \pi( r - r_{\rm in}) }{r_{\rm out} - r_{\rm in}}  \right)   \mbox{    if    }  r \in [r_{\rm in}, r_{\rm out}]\mbox{   , } \nonumber \\
\bar N^2(r) = -10^{-11} \mbox{    otherwise,    }
\end{eqnarray} 
where $r_{\rm in}$ is the radius of the lower radiative--convective interface, 
and $r_{\rm out}$ is the radius of the upper radiative--convective interface. 
The maximum value of $\bar N(r)$ within the radiative zone, 
$N_{\rm rz}$, is one of our input parameters, while that of the convection zones
is merely chosen to be arbitrarily low and has little effect on the outcome 
of the simulation. In what follows, we define the global parameter 
$\sigma_{\star}$ as
\begin{equation}
\sigma_\star = \sqrt{{\rm Pr}} \frac{N_{\rm rz}}{\Omega_\star}\mbox{   . }
\end{equation}
Note that the Prandtl number is assumed to be constant (see below).

In all simulations, the star is assumed to be rotating at the mean angular 
velocity $\Omega_\star = 3\times 10^{-4}$rad/s. In all cases, the applied forcing by 
differential rotation is 
\begin{equation}
\bu_{\rm cz}(r,\theta) = \frac{r \sin\theta \Omega_{\rm cz}(\theta)}{2} \left[2+\tanh\left(\frac{r-r_{\rm out}}{\Delta_{\rm out} }\right) + \tanh\left(\frac{r_{\rm in}-r}{\Delta_{\rm in} }\right)\right] \mbox{  } \ephi \mbox{   , }
\end{equation}
where 
\begin{equation}
\Omega_{\rm cz}(\theta) = \Omega_{\rm eq} \left( 1 - a_2 \cos^2\theta \right)\mbox{   , }
\label{eq:ocz}
\end{equation}
with 
\begin{eqnarray}
&& a_2 = 0.01 \mbox{   and   } \Omega_{\rm eq} = \Omega_\star \left( 1- \frac{3a_2}{15} \right)^{-1} \mbox{   , }
\end{eqnarray}
to ensure that the total applied angular momentum to the system is zero. 
Note that for simplicity, the same differential rotation is chosen in both convection 
zones, and is measured by the parameter $a_2$. This value is chosen to 
be fairly small, since the observed differential rotation 
of rapidly rotating stars is typically quite small. 
In practice, since we are studying a linear
problem, the solutions scale linearly with $a_2$.
Finally, the expression for the non-dimensional quantity $\tau(r)$ is the same as 
that given in (\ref{eq:tau}) with 
$\Lambda_{\rm in} = \Lambda_{\rm out} = 10$ in both inner 
and outer convection zones. 

The total diffusivities (i.e. the sum of the microscopic and turbulent components) 
are assumed to have the following profiles: 
\begin{eqnarray}
&& \bar \kappa(r) + \kappa_{\rm turb}(r) = \kappa_c e^{r/D_\kappa} + \frac{\kappa_{\rm turb}}{2} \left[ 2 + \tanh\left( \frac{r - r_{\rm out}}{\Delta_{\rm out}} \right) + \tanh\left( \frac{r_{\rm in}-r}{\Delta_{\rm in}} \right) \right]  \mbox{  ,} \nonumber \\
&& \bar \nu(r) + \nu_{\rm turb} (r) = {\rm Pr} \, (\bar \kappa(r) + \kappa_{\rm turb}(r)) \mbox{  .} 
\label{eq:kappanucart}
\end{eqnarray}
The selected exponential profile for the microscopic 
part of $\bar \kappa(r)$ is 
not too dissimilar from that of a real star in the 1.3-1.5$M_\odot$ range 
provided $D_\kappa/R_\star = 0.14$. 
We define the inverse Peclet number $E_{\kappa c} = \kappa_c /R_\star^2 \Omega_\star$. 
The value of $E_{\kappa c}$ will be varied in the various simulations, and decreased as 
much as possible to reach the asymptotic stellar regime. 
We take the Prandtl number to be constant and equal to Pr$ = 10^{-2}$. 
The selected value of Pr is chosen to be smaller than one to respect 
stellar conditions, but larger than the actual stellar values 
(which are of order of $10^{-6}$ typically) to ease the numerical computations. 
The ``convective'' value $\kappa_{\rm turb}$ is fixed so that the 
non-dimensional $E_{\kappa,{\rm turb}}$ is equal to $10^{-1}$. 
The overshoot layer depths $\Delta_{\rm in}$ and $\Delta_{\rm out}$ 
are taken to be 0.01. While these choices are fairly arbitrary, 
they are reasonable given our qualitative goals.

The computational domain is a spherical shell with the outer boundary 
located at $0.95 R_\star$ and the inner boundary at 
$0.01 R_\star$. The outer boundary is chosen to be well-below 
the stellar surface to avoid numerical complications related to the very rapidly changing 
background in the region $r > 0.95 R_\star$. The inner boundary is chosen to 
be well-within the inner convection zone, but excludes the origin to avoid
coordinate singularities.
The upper and lower boundaries are assumed to be impermeable and stress-free,
with $\tilde T = 0$. 

The numerical method of solution is based on the expansion of the governing
equations onto the spherical coordinate system, followed
by their projection onto Chebishev polynomials
$T_n(\cos\theta)$, and finally, solution of the resulting ODE system in $r$ 
using a Newton-Raphson-Kantorovich algorithm. The typical solutions 
shown have 3000 meshpoints and 70 Fourier modes. For more detail 
on the numerical algorithm, see Garaud (2001) and Garaud \& Garaud (2008).

\subsection{Typical solution}

\begin{figure}[h!]
\centerline{\epsfig{file=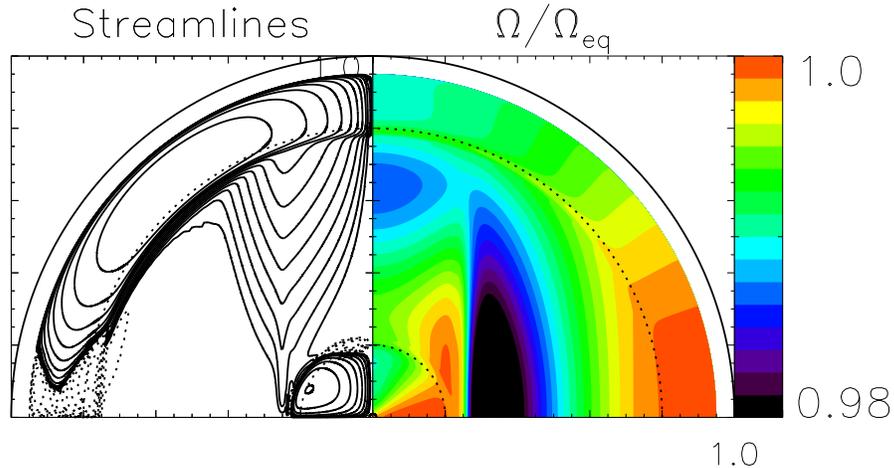,width=16cm}}
\caption{Numerical solution for $r_{\rm out}=0.8R_\star $, 
$r_{\rm in}=0.2R_\star $,
  $E_{\kappa c}= 10^{-7}$ and $\sigma_\star = 0.1$. The two quadrants
  represent the stellar interior, with the radiative--convective
  interfaces denoted as dotted circles. The left-side panel shows
  selected streamlines of the flow, as logarithmically-spaced contours of
  the stream-function. Solid lines represent clockwise flow, while
  dotted lines represent counter-clockwise flow. The right-side panel
  shows the normalized differential rotation profile.}
\label{fig:sample}
\end{figure}

Solutions have been computed for a wide range of values of the parameters 
$d_{\rm in} = r_{\rm in}$ and $d_{\rm out} = R_\star - r_{\rm out}$ (the 
respective depths of the inner and outer convection zones), 
$E_{\kappa c}$ and $\sigma_\star$. For low enough diffusivities 
the overall structure of the radiative zone flows converges to a 
pattern which only depends on $d_{\rm in}$, $d_{\rm out}$ and $\sigma_\star$ (and 
with an amplitude which scales as the calculated $W_{\rm rz}$). 
Figure \ref{fig:sample} shows a representative example of the kind of 
2D flow structure found within the star, for $E_{\kappa c} = 10^{-7}$ and 
$\sigma_\star = 0.1$. An artificial case with fairly thick 
convective zones ($r_{\rm in} = 0.2R_\star $ and $r_{\rm out} = 0.8R_\star$) 
was chosen to make it easier to visualise the results. 

\paragraph {Discussion of the meridional flow structure. }
Representative streamlines are shown in the left-side panel of 
Figure \ref{fig:sample} and reveal 
the structure of the meridional flows driven by gyroscopic pumping. In the 
outer convection zone, we observe a dominant 
cell in mid and high latitudes, poleward near the surface
and equatorward near $r_{\rm out}$. In addition, 
a small counter-cell of fairly surprising shape is observed in the equatorial 
region. The inner convection zone also has a dominant cell of the same 
vorticity (with poleward flows in the outer layers, and a deep equatorward 
return flow), and a small counter-cell located just above it along the polar 
axis. The structure of the two dominant cells in the respective convective 
zones are easily understood from angular momentum balance. The smaller 
counter-cell above the inner convection zone is required to match the 
flows downwelling from the outer convection zone to the inner core flows. 

The bulk of the mass flux generated in a given 
convection zone returns within, or close to the edge of that same 
convection zone. However, weak flows from the polar region of the outer 
convective zone, and from the equatorial region of the inner convection zone
escape and mix the cylinder tangent to the convective core. Note how, by 
contrast, the region outside of the tangent cylinder is mostly quiescent. 

\begin{figure}[h!]
\centerline{\epsfig{file=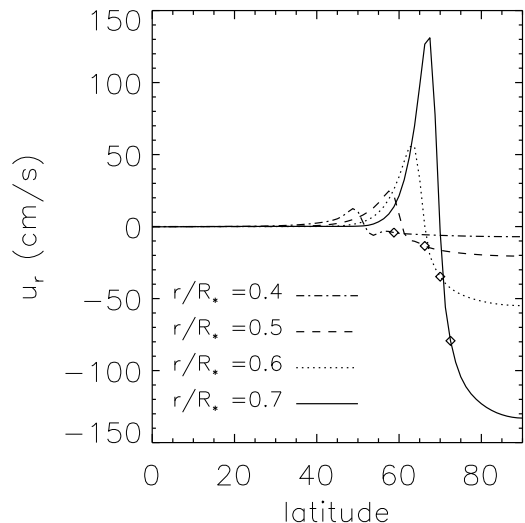,width=8cm}}
\caption{Variation of $u_r$ with latitude for four different non-dimensional
 radii $r/R_\star$ for the
 simulation shown in Figure \ref{fig:sample}. The latitudinal position of the
 tangent cylinder at each radius is marked by a small diamond. 
This figure illustrates the polar
 downwelling, the return flow just outside the tangent cylinder and the mostly
 negligible mixing at low latitudes. Note that the calculated value of $u_r$ 
should not be taken at face value: it needs to be rescaled appropriately 
in order to estimate mixing velocities in real stars (see \S\ref{sec:analytical}). See \S\ref{sec:lidip} for more quantitative estimates in Li-dip stars.}
\label{fig:downwell}
\end{figure}
\begin{figure}[h!]
\centerline{\epsfig{file=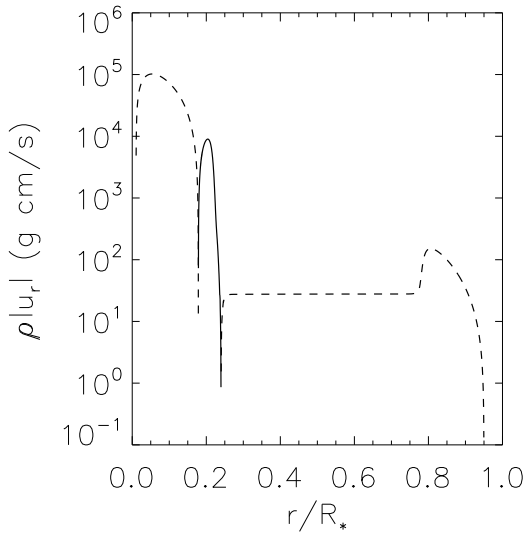,width=8cm}}
\caption{Variation of $\bar \rho |u_r|$ with depth near the polar axis
  (at a latitude of $88^\circ$) in the simulation of Figure
  \ref{fig:sample}. A logarithmic scale is used to visualize
  simultaneously the convection zone flows and the radiative zone
  flows. The solid line denotes upwelling flows, while the dashed line
  denotes downwelling flows. Note the radiative region $r/R_\star \in
  [0.2,0.8]$  where $\bar \rho u_r$ is constant as expected from the
  analytical solution. }
\label{fig:rhour}
\end{figure}

The variation of the vertical flow amplitude with latitude for the 
simulation of Figure \ref{fig:sample} can be seen more clearly in Figure 
\ref{fig:downwell}: it illustrates the downward pumping 
of the flows in the polar regions, while mixing in the equatorial region
is mostly negligible. The variation of the amplitude of the flows with depth 
is well-explained by the variation in the 
background density: flows pumped downward along the polar axis
roughly satisfy $u_r = W_{\rm rz}/\bar \rho $ 
(where $u_r$ is the radial velocity) where $W_{\rm rz}$ is constant. 
This constraint is more easily seen in Figure \ref{fig:rhour} 
which shows the profile of $\bar\rho u_r$ at $88^\circ$ latitude for 
the simulation shown in Figure \ref{fig:sample}. Within the radiative zone 
we see that $\bar \rho u_r$ is roughly constant as expected. Figure \ref{fig:rhour} also illustrates how only a small fraction of the pumped mass flux 
enters the radiative zone, while most of it remains within the generating convection zone.

\paragraph{Discussion of the azimuthal flow structure.}
As seen in Figure \ref{fig:sample}, both convection zones exhibit a rotation
profile close to the imposed profile (i.e. with $\Omega/\Omega_{\rm eq}$ 
ranging from 1-$a_2$ to 1 between the pole and the equator, 
and with $\Omega$ nearly constant with radius), 
as expected. Meanwhile, the radiative zone exhibits a similar
level of differential rotation, with a striking shear layer near 
the tangent cylinder. This shear is presumably caused by the 
deposition of negative angular momentum by the meridional flows
as they carry fluid away from the rotation axis and begin to rise up 
in the radiative interior again. Other numerical simulations (not shown here) 
show that this feature is stronger when flows are stronger, 
i.e. when the system is more weakly 
thermally stratified or more rapidly rotating ($\sigma_\star$ smaller). An 
interesting consequence of this shear layer, however, is the possibility that 
it may become unstable to Rayleigh instabilities in the vicinity of the 
tangent cylinder, where for extreme cases the specific angular momentum 
may locally begin to decrease with distance from the polar axis.

\subsection{Comparison between Cartesian model predictions and spherical 
model solutions}
\label{sec:sphertocart}

The results presented in the previous section show that, for stars with 
$\sigma_\star < 1$, the downward pumped mass flux into the radiative 
zone $W_{\rm rz}$ is indeed constant with depth along the rotation axis,
as found in the Cartesian model analysis of \S\ref{sec:cart}.
We can now compare more quantitatively 
the results of these spherical simulations with equivalent 
Cartesian solutions for $W_{\rm rz}$, first in order to verify the scalings
derived in (\ref{eq:wrzconst}) and then to see if there exists a simple 
relationship between the Cartesian model velocities and the 
spherical model velocities.

We first run a series of two-dimensional spherical simulations, based on the 
model presented in \S\ref{sec:sphermod}, with the following parameters varied:
\begin{itemize}
\item We consider three different geometries: 
$(r_{\rm in}/R_\star = 0.05,\, r_{\rm out}/R_\star =0.9)$
$(r_{\rm in}/R_\star = 0.1,\, r_{\rm out}/R_\star =0.8)$, and 
$(r_{\rm in}/R_\star = 0.2,\, r_{\rm out}/R_\star =0.8)$. The purpose is 
to explore the effect of varying the convection zone sizes on the predicted
velocities.
\item For the case with $r_{\rm in}/R_\star = 0.1, r_{\rm out}/R_\star =0.8$
we consider two different values of $\sigma_\star$ (by changing $N_{\rm rz}$): 
$\sigma_\star = 0.1$ and $\sigma_\star = 0.03$. For the other two geometries 
$\sigma_\star $ is fixed to be $0.1$. 
\item Finally, we consider a range of inverse Peclet numbers 
from $10^{-5}$ down to the lowest achievable value, $E_{\kappa c} = 10^{-10}$. 
\end{itemize}
In all cases, we measure the mass flux $W_{\rm rz}^{\rm spher} = \bar \rho |u_r|$ at $r/R_\star = 0.5$, 
at a latitude of $85^\circ$. Note that this choice is fairly arbitrary: 
$\bar \rho |u_r|$ does not change with depth nor with latitude much as long
as the point selected lies well-within the tangent cylinder. 

In order to compare the value obtained in the spherical case with 
Cartesian model simulations, 
we integrate the equivalent equations and boundary conditions, 
now expressed in a Cartesian coordinate system\footnote{Note that these
equations are different from the ones used in \S\ref{sec:cart} 
and are indeed the Cartesian expression of (\ref{eq:gov_eqs}). 
Specifically, they differ from those of \ref{sec:cart}  by using 
the correct viscous stress tensor, the correct heat flux, and the 
full linearized equation of state.} using exactly the same geometries 
($z_{\rm in} = r_{\rm in}/R_\star$, $z_{\rm out} = r_{\rm out}/R_\star$), 
background profiles, diffusivities 
and convective turnover timescale 
as in the spherical cases described above (e.g. equation (\ref{eq:tau}), 
and equation (\ref{eq:kappanucart}) with $r$ replaced by $z$). We construct the forcing 
velocity $\hat u_{\rm cz}(z) e^{iky}$ based on the differential rotation profile
$\Omega_{\rm cz}(r,\theta)$ in the following way: we take $k=2$, and 
set
\begin{equation}
\hat u_{\rm cz}(z) = \frac{ a_2}{2} \left[2+\tanh\left(\frac{z-z_{\rm out}}{\Delta_{\rm out} }\right) + \tanh\left(\frac{z_{\rm in}-z}{\Delta_{\rm in} }\right)\right] \mbox{  ,}
\end{equation}
in both inner and outer convection zones. We then measure 
$W_{\rm rz}^{\rm cart} =\bar \rho \hat w $ at the same height ($z=0.5$).

\begin{figure}[h!]
\centerline{\epsfig{file=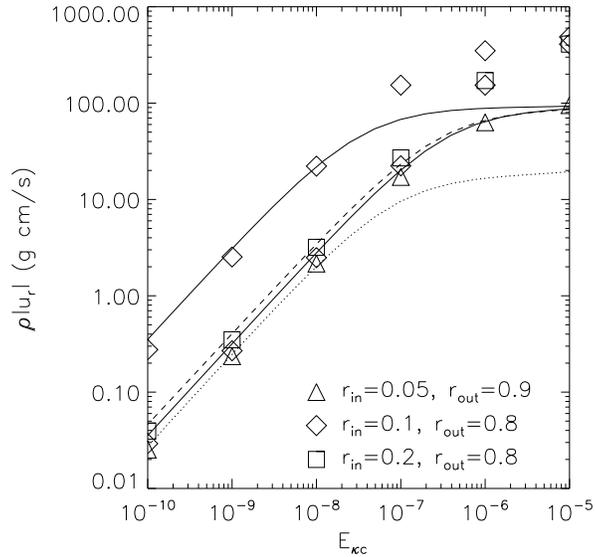,width=8cm}}
\caption{Comparison between $\bar \rho |u_{r}|$ extracted from the 
spherical model calculations (symbols) as described in the main text, and 
the equivalent Cartesian model calculation (lines) where 
$W^{\rm cart}_{\rm rz}$ is divided
by a factor of 2$\pi$. The solid lines correspond to the case
where $r_{\rm in} = 0.1 R_\star$ and $r_{\rm out} = 0.8$, for $\sigma_\star = 0.1$ 
and $\sigma_\star = 0.03$ for the 
lower and upper line respectively. The dashed line corresponds 
to $r_{\rm in} = 0.2 R_\star$ and $r_{\rm out} = 0.8 R_\star$, 
while the dotted line corresponds to $r_{\rm in} = 0.1 R_\star$ 
and $r_{\rm out} = 0.9 R_\star$. }
\label{fig:compare_spher}
\end{figure}

As shown in Figure \ref{fig:compare_spher} we find that there is an excellent
agreement of the Cartesian model calculations with the spherical model calculations, 
for {\it all} sets of simulations at low enough values of the diffusivities, 
provided we divide the Cartesian model results by a factor of 2$\pi$:
\begin{equation}
W_{\rm rz}^{\rm spher} = \frac{W_{\rm rz}^{\rm cart}}{2\pi} \mbox{  .}
\label{eq:cart2spher}
\end{equation}
The discrepancy for higher values of the diffusivities appears to 
arise when the diffusive layer thicknesses 
become of the order of the modelled 
structures (i.e. the thickness of the outer convection zone or the 
width of the tangent cylinder).

This result implies that the overall scalings derived are indeed correct, 
but furthermore that it is possible to use the simplified Cartesian model
to get {\it very precise} estimates of the mass flux into the radiative 
zone within the tangent cylinder. The good fit between the two sets of 
simulations can presumably
be attributed to the fact that as long as $r_{\rm in} \ll R_\star$, 
the tangent cylinder is  
quite thin and curvature effects should indeed be negligible. 
The multiplicative factor 
of 1/2$\pi$ is not obvious a priori (hence the need for this exercise), but is 
not particularly surprising either. We attribute it
to the fact that in the cylindrical case, the rotation axis is ``infinitely far'' away from 
the region where the flows are calculated, whereas it plays an important role in the 
calculation in the spherical case.

\subsection{Summary}

To summarize this section, we have seen that the gyroscopic pumping 
mechanism still works (as expected!) in a spherical geometry, and that the 
dynamics within the cylinder tangent to the inner convective core (for stars
with two convective zones, see Figure \ref{fig:gyro2}) 
are very similar, qualitatively and quantitatively, 
to the dynamics studied in \S\ref{sec:cart}. In particular, we 
found that the pumped mass flux within this tangent cylinder, as measured 
in full two-dimensional spherical geometry calculations, is equal to the
equivalent pumped flux calculated in the Cartesian case, but divided
by a factor of $2\pi$. 

This important result provides an interesting and practical mean of 
getting precise estimates for the rate of mixing induced by gyroscopic 
pumping, which can in principle be used in stellar evolution models. 
In the following section, we provide an example of application of this idea, 
to the Li-dip problem.

\section{Application to the Li dip problem} 
\label{sec:lidip}

\subsection{Introduction}
\label{sec:liintro}

 An outstanding problem in stellar astrophysics concerns
 the measured abundance of the rare light element Lithium in the
 atmospheres of F-type main-sequence stars (for reviews see Boesgaard
2005 and Anthony-Twarog et al. 2009).
 Lithium burns by nuclear reactions at
 temperatures of 2.5 $\times 10^6$ K or above in these stars,  but the
 depletion is not evident at the surface unless there is a
 mixing mechanism to bring the Li down to layers at that temperature
 at some time during the evolution of the star. In the mass
 range of interest, between 1.1--1.6 M$_\odot$, the pre-main-
 sequence convection zone does not extend down to high enough 
 temperatures to result in
 appreciable Li depletion at the surface, and in fact most
 main sequence Pop I  stars in this mass range have Li
 abundances ($N_{\rm Li} \approx 2 \times 10^{-9}$ that of hydrogen by number)
 characteristic of those in the youngest stars,
 indicating little, if any depletion. 

 However, as first 
 discovered in the Hyades cluster (Boesgaard \& Tripico 1986)
 there is a narrow range in effective temperature 6400 K $< T_{\rm eff} < 6900$ K
 (spectral types F6--F0) where a sharp dip 
 in the Li abundance is observed, 
with a minimum value of $ N_{\rm Li} <10^{-11}$ at $T_{\rm eff} 
 \approx 6650$ K. The mass range within the dip is 1.3--1.5 M$_\odot$.
 The depth of the surface convection zone decreases rapidly as 
 $T_{\rm eff}$ increases across the dip.
 The same temperature range also corresponds to a rapid change in 
 spectroscopic rotational velocities (Boesgaard 1987; Wolff \& Simon 1997),
 with the stars
 around $T_{\rm eff} = 7000$ K rotating with $v \sin i$ up to 150 km/s
 and those at $T_{\rm eff} = 6400$ K with only 20 km/s.
 A similar  dip is also observed in the Praesepe cluster 
 (Soderblom et al. 1993a)
 with an age similar to that of the Hyades (600-700 Myr). In 
 the much younger Pleiades cluster (100 Myr) the dip  at about
 $T_{\rm eff} = 6700$ K is marginal
(Soderblom et al. 1993b) or not present (Boesgaard 2005). In the even
 younger cluster $\alpha$ Per (50 Myr) the dip is also not yet evident 
 (Balachandran et al. 1996). Nevertheless the data suggest that at
least some Li depletion occurs relatively early, before an age of 200 Myr
(Anthony-Twarog et al. 2009). In the Hyades, a similar dip, but not as
 deep, is observed for the light element beryllium (Boesgaard \& King
 2002). In the older cluster IC 4651 (1--2 Gyr) both the Li dip
 and the (less deep) Be dip are observed (Smiljanic et al. 2010).
For further details on the properties of
 the dip in various clusters  see Pinsonneault (1997) and Anthony-Twarog
 et al. (2009).

 Main-sequence surface convection zones in this mass range do not
 extend deep enough to mix Li and Be down to their respective burning radii, 
 so the challenge
 is to find another mixing process that operates only in this
 particular range of spectral types.  As summarized by Pinsonneault (1997)
 and  Anthony-Twarog et al (2009), the various proposed
 mixing mechanisms to explain the dip can be divided roughly into
 three types: mass loss, diffusion, or slow mixing as a consequence of
 rotation or waves. Schramm et al. (1990) suggest that the temperature
 range of the lithium dip also corresponds to that of the pulsational
 instability strip where it intersects the main sequence, so that a
 slow mass loss rate, induced by low-amplitude pulsations,  could
 simply remove the lithium remaining in the surface layers.

 Michaud (1986; see also Richer \& Michaud 1993) explained the dip by
 diffusion and gravitational settling  of Li atoms out the
 bottom of the convection zone. In their model the
 cool side of the dip arises from the increasing effectiveness
 of diffusion once the convection zone becomes thin, and the hot
 side is explained by radiative upward acceleration which counteracts
 the diffusion once the star becomes hot enough. To obtain good agreement
 with observations, a small amount of mass loss is also required in
 the theory. However, diffusion models tend to deplete Li and Be 
 at about the same rate, and are not consistent with observations.
 
 Some form of rotationally-induced mixing seems to be the most
 promising effect to explain the observations (Pinsonneault 1997),
 in particular the Li/Be ratio in the dip (Deliyannis \& Pinsonneault
 1997).  This process can circulate Li out of the surface convection
 zone down to layers where it can be destroyed, but no entirely
 satisfactory model has yet been found. Such mixing can be induced
 by gravity waves  generated by the surface convection zone (Garcia Lopez
\& Spruit 1991, Talon \&
 Charbonnel 2003) or by meridional circulation or secular shear instabilities
 (Deliyannis \& Pinsonneault 1997). In the rotational mixing models
 the cool side of the dip is explained by the increase in rotational
 velocity as $T_{\rm eff}$ increases, possibly combined with the
gravity wave model. The hot side is much more
 difficult to explain; Talon \& Charbonnel (1998) propose a model
involving wind-driven meridional circulation and turbulent transport
induced by differential rotation, based on earlier work by Zahn (1992) and
Talon \& Zahn (1997). This type of model was shown to be consistent
with both Li and Be observations around the gap in  IC 4651 (Smiljanic
et al. 2010).

\subsection{Li and Be depletion by gyroscopic pumping}

In this section, we are primarily interested in determining 
the effect of gyroscopic pumping on Li and Be depletion
as a stand-alone mechanism (i.e. in the absence of any of the 
effects described in \S\ref{sec:liintro}). 
We consider stars in the mass range of the Li dip, namely 
$1.3M_\odot-1.6M_\odot$. We use the results
of \S\ref{sec:spher} to estimate the depletion rate
of Li and Be in the surface layers of these stars, induced
by gyroscopic pumping, as follows. First, we find numerical estimates 
for the pumped mass flux out of the convective envelope and 
flowing into the deep interior within the cylinder tangent
to the inner core (see Figure \ref{fig:gyro2} 
and also Figure \ref{fig:gyrofinal}).
In order to do this, we solve the set of equations (\ref{eq:gov_eqs}) expressed
in a Cartesian geometry using a real stellar background model,
extract the desired value of the pumped mass flux $W_{\rm rz}^{\rm cart}$, 
and then use the rule (\ref{eq:cart2spher}) to estimate 
the equivalent pumped mass flux in the more realistic case of 
a spherical star. 
Using simple geometrical arguments, we then construct and solve
evolution equations for the surface Li and Be abundances, which 
can be compared with observations.

\subsubsection{Background model}

We use the code developed by Bodenheimer 
et al. (2007) to construct a sequence of reference 
background models in the range $1.3M_\odot- 1.6M_\odot$, 
evolved from the ZAMS up to 
300 Myr which is about half the age of the Hyades cluster. 
For reference, this code solves the standard equations of stellar
structure and evolution, and uses a gray model atmosphere as an outer 
boundary condition. We assume a solar initial composition. The code is 
calibrated to match the Sun's observed properties at 4.57 Gyr. 

Table 1 summarizes various properties 
of these modelled stars at age 300 Myr:
the stellar radius $R_\star$, the respective 
depths of the inner and outer convection zones $d_{\rm in}$ and $d_{\rm out}$,
an estimate of the convective velocities in the bulk of each 
convection zone, $v_{\rm conv}^{\rm in}$ and $v_{\rm conv}^{\rm out}$
and finally the radii $r_{\rm Li}$ and $r_{\rm Be}$ below which the local
stellar temperature exceeds the Li-burning and Be-burning temperatures 
of $2.5 \times 10^6$ K and $3.5 \times 10^6$ K respectively. 

\begin{table}
\vspace{0.05cm} 
\begin{tabular}{|p{1cm}|p{1cm}|p{2cm}|p{2cm}|p{1.5cm}|p{1.5cm}|p{1cm}|p{1cm}|p{2cm}|}
\hline
$ M_\star $ & $R_\star$ & $d_{\rm in}$ & $d_{\rm out}$ & $v_{\rm conv}^{\rm in}$ & $v_{\rm conv}^{\rm out}$ & $r_{\rm Li}$ & $r_{\rm Be}$ & $\Omega_\star$ \\
\hline
1.30 & 8.84  & $5.5\times 10^{-2}$ & $1.2\times 10^{-1}$ & $ 8.0 \times 10^2$ &  $2 \times 10^4 $ & 0.61 & 0.51& $8.8 \times 10^{-5}$   \\
1.35 & 9.28  & $5.8 \times 10^{-2}$ & $1.0 \times 10^{-1}$ & $ 5.0 \times 10^2$ &  $1 \times 10^4 $ & 0.60 & 0.50 & $1.0 \times 10^{-4}$ \\
1.40 & 9.69  & $6.3 \times 10^{-2}$ & $7.5 \times 10^{-2}$ & $ 1.0 \times 10^3$ &  $3 \times 10^4 $ & 0.59 & 0.49 & $1.3 \times 10^{-4}$ \\
1.45 & 10.1  & $6.9 \times 10^{-2}$ & $5.3 \times 10^{-2}$ & $ 1.2 \times 10^3$ &  $5 \times 10^4 $ & 0.57 & 0.48 & $1.5 \times 10^{-4}$\\
1.50 & 10.3 & $7.5 \times 10^{-2}$ & $2.9 \times 10^{-2}$ & $ 1.5 \times 10^3$ &  $1 \times 10^5 $  & 0.57 & 0.48  & $1.5 \times 10^{-4}$\\
1.55 & 10.6  & $8.0 \times 10^{-2}$ & $1.4 \times 10^{-2}$ & $ 1.8 \times 10^3$ &  $1 \times 10^5 $ & 0.57 & 0.48 & $1.5 \times 10^{-4}$\\
1.60 & 10.8  & $8.5 \times 10^{-2}$ & $6.5 \times 10^{-3}$ & $ 2.0 \times 10^3$ &  $1 \times 10^5 $  & 0.57 & 0.48 & $1.5 \times 10^{-4}$ \\
\hline
\end{tabular}
\caption{Stellar features as extracted from our reference $1.3M_\odot- 1.6M_\odot$
background models evolved to 300 Myr. From left to right: the stellar mass (in units of $M_\odot$), the
stellar radius (in units of $10^{10}$cm), the depths of the inner and outer convection zones in units
of the stellar radius, the convective velocities in each convection zone in cm/s, the Li and Be
burning radii (see main text) in units of the stellar radius, and finally an estimated
rotation rate for the star (see main text) in radians per second. }
\end{table}

These reference stars are assumed to rotate with an angular velocity 
$\Omega_\star$ derived from the results of Wolff \& Simon (1997), who
provide estimates for the mean $v \sin i$ of stars in various mass ranges 
and ages (see their Table 4). We first note that in the mass range considered,
the mean rotational velocities do not change much between the Pleiades age 
and the Hyades age. We interpolate the observations of these two clusters
to an age of approximately 300 Myr. 
 We also note that at the Pleiades age their rather high $1.4-1.5M_\odot$ 
measurement only has 4 data points; we discard it. 
We then interpolate their remaining results
to our selected stellar masses to get:
$<v \sin i>_{\rm obs}(1.3M_\odot) = 50$km/s, 
$<v \sin i>_{\rm obs}(1.35M_\odot) = 61$km/s, 
$<v \sin i>_{\rm obs}(1.4M_\odot) = 79$km/s,  
$<v \sin i>_{\rm obs}(1.45M_\odot) = 96$km/s, 
$<v \sin i>_{\rm obs}(1.5M_\odot) = 98$km/s,
$<v \sin i>_{\rm obs}(1.55M_\odot) = 100$km/s and  
$<v \sin i>_{\rm obs}(1.6M_\odot) = 100$km/s.
To convert the mean $v \sin i$ measurements 
into rotational velocities, we note that the mean value of $\sin i$ over
all possible measurements is (very roughly) $2/\pi$. In that case, we take
\begin{equation}
\Omega_\star = \frac{ \pi <v \sin i>_{\rm obs}}{2R_\star}\mbox{  .} 
\end{equation}
The resulting $\Omega_\star$ values are listed in Table 1. Note that 
because of the change in the stellar radius across the selected mass
range, $\Omega_\star$ does not vary too much.

We also extract from the stellar models 
the radial density, temperature, opacity and buoyancy frequency 
profiles within the stars, which are used as the background state
for the Cartesian calculation of the gyroscopically
pumped mass flux. From these quantities, we continue constructing the model as follows. 
The microscopic viscosity and thermal diffusivity profiles 
$\bar \nu(r)$ and $\bar \kappa(r)$ are calculated 
using the formula given by Gough (2007) (see also Garaud \& Garaud 2008). 
The turbulent diffusivities in the convective regions 
(see equation (\ref{eq:eturbs}))
are derived from the model convective velocities as:
\begin{equation}
E^{\rm in}_{\kappa,{\rm turb}}(z) =  E^{\rm in}_{\nu,{\rm turb}}(z) = \frac{v^{\rm in}_{\rm conv} d_{\rm in}}{R_\star^2\Omega_\star} \mbox{  , } 
\end{equation}
in the convective core, and similarly for the outer convection zone. 
The overshoot depths $\Delta_{\rm in}$ and $\Delta_{\rm out}$ may be different near the 
inner and outer convective regions, 
and are taken to be equal to 10\% of the local
pressure scaleheight at the respective radiative--convective interfaces.  

The convection zones are both assumed 
to be rotating differentially with the profile given in equation 
(\ref{eq:ocz}). The
parameter $a_2$, which to a good approximation is equal to the 
difference between the equator and the polar rotation rate, normalized
by $\Omega_\star$, is assumed for simplicity to be the same in both convection 
zones, and is a free parameter in the problem (recall that the predicted
velocities scale linearly with $a_2$).
The convective velocities are used to derive the quantities 
$\Lambda_{\rm in}$ and $\Lambda_{\rm out}$ which characterize the relaxation
timescale to this assumed differential rotation profile 
(see equation (\ref{eq:tau})). We take 
\begin{equation}
\Lambda_{\rm in} = \frac{v_{\rm conv}^{\rm in}}{\Omega_\star R_\star} \mbox{  , } 
\end{equation}
and similarly for $\Lambda_{\rm out}$. Note that this quantity is the most
difficult to relate to real stellar parameters, since it 
refers to our very simplified parametrization of the effects of turbulence.
Here we have used $R_\star$ as a typical lengthscale instead of the
depth of the convective region in calculating the relaxation timescale. 
The reasoning behind this 
choice is that angular momentum has to be transported all the way from the 
pole to the equator (i.e. a typical lengthscale $R_\star$) 
for the large-scale differential rotation profile to be established.

Finally, it is crucial to note that while some of our choices (of $\Omega_\star$, 
of $\Lambda_{\rm in}$ and $\Lambda_{\rm out}$) are 
arguably arbitrary, it so happens that they do not influence the resulting 
depletion rates much. The 
reasons for this will be explained in detail in \S\ref{sec:lidisc}. 

\subsubsection{Calculated mass flux and depletion timescale}
\label{sec:lipred1}

\begin{figure}[h]
\centerline{\epsfig{file=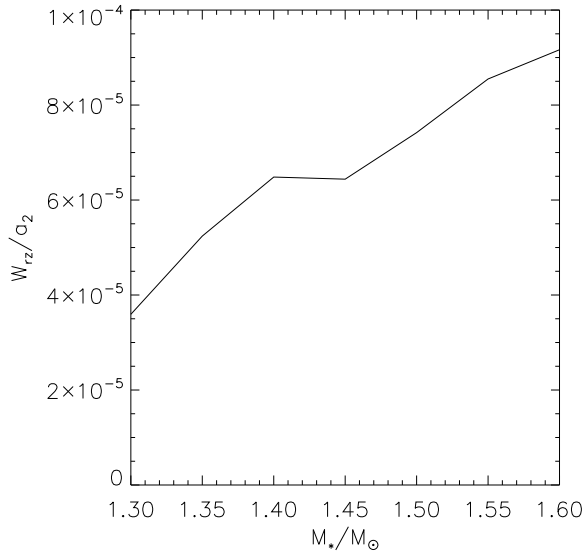,width=8cm}}
\caption{Calculated mass flux $W^{\rm spher}_{\rm rz}$ in the model simulations 
as a function of stellar mass. Note that $W^{\rm spher}_{\rm rz}$ scales 
linearly with $a_2$, hence the rescaling on the $y-$axis. To estimate 
a radial velocity at radius $r$, divide $W_{\rm rz}^{\rm spher}$ by the density at that radius, $\bar \rho(r)$. }
\label{fig:wrzs}
\end{figure}

In order to evaluate the mass flux induced by gyroscopic pumping 
we now integrate the two-point boundary value problem 
(\ref{eq:gov_eqs}), expanded in Cartesian geometry,
with the realistic background 
stellar model described in the previous section. 
The advantage of the Cartesian calculation is that it can indeed be performed
with true stellar values of the diffusivities, using 100,000 meshpoints. 
This would not be possible with spherical-geometry calculations.
We extract from the numerical solutions the quantity 
$W^{\rm cart}_{\rm rz}$ by measuring the value of $\bar \rho \hat w$ 
at $z=0.5$, although, as discussed in \S\ref{sec:cart}, the exact 
height does not matter since $\bar \rho \hat w$ is constant within 
the radiative zone. We first convert this value back into dimensional form,
and then to the spherical case using 
(\ref{eq:cart2spher}) to obtain an estimate for $W^{\rm spher}_{\rm rz}$, 
the local mass flux flowing in the tangent cylinder. The results are 
shown in Figure \ref{fig:wrzs}, and discussed in \S\ref{sec:lidisc}.

\begin{figure}[h]
\centerline{\epsfig{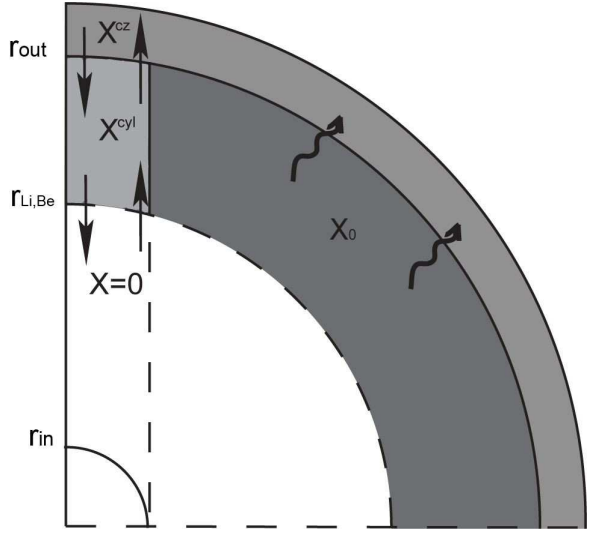}}
\caption{Schematic of the Li (or Be) mass flux in the various regions of a
Li-dip star's interior. The surface and the two radiative--convective interfaces
are shown as solid circles. The dashed line represents the Li-burning (or alternatively the 
Be-burning) radius -- only one of the two is shown on this Figure for clarity.
Fluid is pumped from the surface downward along the rotation axis, and 
returns to the outer convection zone in a thin layer 
close to the edge of the tangent cylinder (vertical dashed line). By mass conservation, both fluxes are the same and equal to $W_{\rm rz}^{\rm spher} \pi r_{\rm in}^2$. The mass abundance of Li (and Be) in the outer convection zone is $X^{\rm cz}$, that in the cylinder $C_{\rm Li,Be}$ is $X^{\rm cyl}$ while $X=0$ below the burning radius. In the region below the outer convection zone but above the burning radius, the abundance ($ X_0$) is essentially primordial. Diffusion (squiggly arrows) brings Li and Be back into the outer convection zone. }
\label{fig:gyrofinal}
\end{figure}

Based on the numerical results from \S\ref{sec:sphermod}, and 
best illustrated in Figure \ref{fig:sample}, we now
approximate the Li circulation pattern in the manner depicted 
in Figure \ref{fig:gyrofinal} (see also Figure \ref{fig:gyro2}). 
Li-rich fluid is pumped down from the outer convective zone, through the 
cylindrical region $C_{\rm Li}$ (delimited in radius by the cylinder 
tangent to the convective core and in the vertical direction by 
$r_{\rm Li}$ and $r_{\rm out}$ respectively), 
down to the Li-burning interior. By mass conservation, 
Li-free material is pumped up from the deep interior, 
through $C_{\rm Li}$, to the outer convective zone.
As a result, $C_{\rm Li}$ and the 
outer convection zone are both progressively depleted in Li with time. 
 
\begin{figure}[h]
\centerline{\epsfig{file=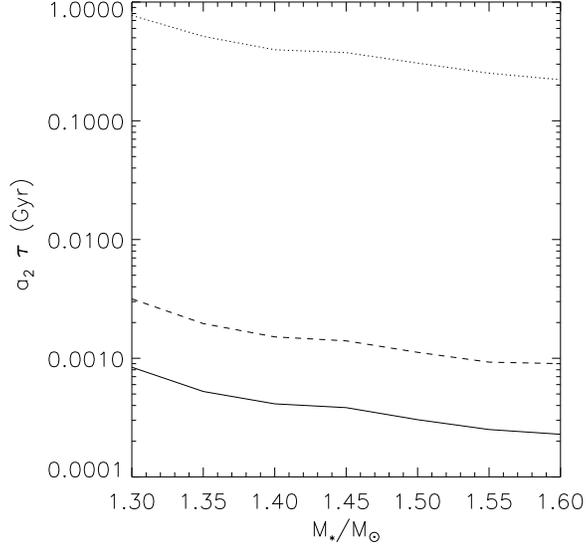,width=8cm}}
\caption{Timescales for the circulation of material from the base of
  the outer  convective zone down to the Li-burning radius (solid
  line), the Be-burning radius (dashed line) and the inner convective
  core (dotted line). Note that since $W^{\rm spher}_{\rm rz}$ scales
  linearly with $a_2$, so do these timescales, hence the rescaling of
  the $y-$axis. For reasonable values of $a_2$ (of the order of a
  percent), the Li- and Be-circulation timescales are smaller than the
  age of the star, while the circulation from the surface down to the
  inner core is of tens of Gyr.}
\label{fig:timescales}
\end{figure}

Note that the timescale for material to flow vertically from the Li-burning radius 
$r_{\rm Li}$ to the base of the outer convection zone $r_{\rm out}$ 
(or vice-versa) can easily be calculated by integrating the inverse of 
the radial velocity $W^{\rm spher}_{\rm rz}/\bar \rho$ between $r_{\rm Li}$ 
and $r_{\rm out}$.
The results, using $W^{\rm spher}_{\rm rz}$ estimated above and the 
true stellar density profiles, 
are shown in Figure \ref{fig:timescales}, 
together with the equivalent timescale for the 
transport of Be from the Be-burning radius $r_{\rm Be}$ 
to the outer convection zone.
In all stars considered, the transport timescales are of the order of 
30-100 Myr and 100-300 Myr for Li and Be 
respectively when the differential rotation rate $a_2$ is one percent.
Note that the transport timescale is not equal to the depletion timescale, 
although the two are related (see Appendix B for detail).  
Figure \ref{fig:timescales} also shows the similarly
calculated timescale required for material to move from 
the convective core to 
the convective envelope. It is interesting 
and reassuring to note that for reasonable values of $a_2$, this timescale 
is of the order of tens of Gyr, so that one would not 
expect to see a strong modification 
of the surface abundances of He or CNO nuclei over the 
lifetime of these stars. This is an important self-consistency check of the model.

We now evaluate the rate of change of the mass fraction of Li in the 
outer convective zone, denoted as $X_{\rm Li}^{\rm cz}$. 
By definition $X_{\rm Li}^{\rm cz} = M_{\rm Li}^{\rm cz}/M^{\rm out}_{\rm cz}$, 
the ratio of the total mass of Li in the outer convection zone to 
the total mass of the outer convection zone. 
Since material flows through $C_{\rm Li}$ on its way up and down, 
we also need to define 
the mass fraction of Li in $C_{\rm Li}$,  $X_{\rm Li}^{\rm cyl} = M_{\rm Li}^{\rm cyl}/M(C_{\rm Li})$, i.e. the total mass of Li
contained in $C_{\rm Li}$ divided by the mass of that region. Note that 
\begin{eqnarray}
M(C_{\rm Li}) &=& \int_{r_{\rm Li}}^{r_{\rm out}} {\rm d}r \int_0^{\sin^{-1}(r_{\rm in}/r)}  \quad 2\pi \bar \rho(r) r^2 \sin\theta {\rm d}\theta\nonumber \\ &=& \int_{r_{\rm Li}}^{r_{\rm out}}   2\pi \bar \rho(r) r^2 \left ( 1 - \sqrt{ 1-\frac{r_{\rm in}^2}{r^2}} \right){\rm d}r \mbox{  ,} 
\label{eq:mli}
\end{eqnarray}
and can easily be integrated numerically. 

Based on Figure \ref{fig:gyrofinal}, we see that
\begin{eqnarray}
&& \dot{M}^{\rm cz}_{\rm Li} =  - \pi r_{\rm in}^2 W^{\rm spher}_{\rm rz}  X^{\rm cz}_{\rm Li} + \pi r_{\rm in}^2 W^{\rm spher}_{\rm rz} X^{\rm cyl}_{\rm Li} \mbox{  ,}  \nonumber \\
&& \dot{M}^{\rm cyl}_{\rm Li} =  \pi r_{\rm in}^2 W^{\rm spher}_{\rm rz}  X^{\rm cz}_{\rm Li} - 2 \pi r_{\rm in}^2 W^{\rm spher}_{\rm rz}  X^{\rm cyl}_{\rm Li}  + 0 \mbox{  ,} 
\label{eq:mlicz1}
\end{eqnarray}
since the total mass flux both up and down, within the tangent cylinder, is equal to $\pi r_{\rm in}^2 W^{\rm spher}_{\rm rz}$. Dividing
the first equation by $M_{\rm cz}^{\rm out}$ and the second by $M(C_{\rm Li})$, we finally get
\begin{eqnarray}
&& \dot{X}^{\rm cz}_{\rm Li} = -\frac{\pi r_{\rm in}^2 W^{\rm spher}_{\rm rz}}{M^{\rm out}_{\rm cz}}  X^{\rm cz}_{\rm Li} + \frac{\pi r_{\rm in}^2 W^{\rm spher}_{\rm rz}}{M^{\rm out}_{\rm cz}}  X^{\rm cyl}_{\rm Li} \mbox{  ,}  \nonumber \\
&& \dot{X}^{\rm cyl}_{\rm Li} = \frac{\pi r_{\rm in}^2 W^{\rm spher}_{\rm rz}}{M(C_{\rm Li})}  X^{\rm cz}_{\rm Li} - \frac{2\pi r_{\rm in}^2 W^{\rm spher}_{\rm rz}}{M(C_{\rm Li})}  X^{\rm cyl}_{\rm Li}  \mbox{  ,} 
\label{eq:mlidot1}
\end{eqnarray}
which implicitly defines two timescales,
\begin{eqnarray}
&& \tau^{\rm cz}_{\rm pump} = \frac {M^{\rm out}_{\rm cz}}{\pi r_{\rm in}^2 W^{\rm spher}_{\rm rz}} \mbox{  ,} \nonumber \\
&& \tau^{\rm cyl}_{\rm Li,pump} = \frac {M(C_{\rm Li})}{\pi r_{\rm in}^2 W^{\rm spher}_{\rm rz}}\mbox{  .} 
\end{eqnarray}
The timescale $\tau^{\rm cz}_{\rm pump}$ is the characteristic timescale over which the 
material within the outer convection zone is recirculated by gyroscopic pumping, 
while $\tau^{\rm cyl}_{\rm Li, pump}$ is the timescale over which the material within the 
cylinder $C_{\rm Li}$ is recirculated. If $\tau^{\rm cz}_{\rm pump} \ll \tau^{\rm cyl}_{\rm Li, pump}$
then $X^{\rm cz}_{\rm Li} \simeq X^{\rm cyl}_{\rm Li}$ at all times. Meanwhile, 
if $\tau^{\rm cz}_{\rm pump}$ is of the order of or greater than 
$\tau^{\rm cyl}_{\rm Li, pump}$, then $X^{\rm cz}_{\rm Li}$ can 
differ from $X^{\rm cyl}_{\rm Li}$ significantly. 

The set of equations (\ref{eq:mlidot1}) can easily be solved analytically given the initial 
condition $X^{\rm cz}_{\rm Li}(t=0) = X^{\rm cyl}_{\rm Li}(t=0) = X_{\rm Li,0}$, 
the initial Li abundance in the star. The calculation is detailed in Appendix B. 
A very similar calculation can be done to estimate the surface Be 
abundance $X^{\rm cz}_{\rm Be}$ 
as a function of time. The only difference is that one should replace $r_{\rm Li}$ with 
$r_{\rm Be}$, the Be-burning radius, which defines the slightly larger 
cylindrical region $C_{\rm Be}$, and 
equivalently the quantity  $X^{\rm cyl}_{\rm Be}$. The timescale $\tau^{\rm cz}_{\rm pump}$ 
remains the same, while we define
\begin{equation}
\tau^{\rm cyl}_{\rm Be,pump} = \frac {M(C_{\rm Be})}{\pi r_{\rm in}^2 W^{\rm spher}_{\rm rz}} \mbox{ .}
\end{equation}

\begin{figure}[h]
\centerline{\epsfig{file=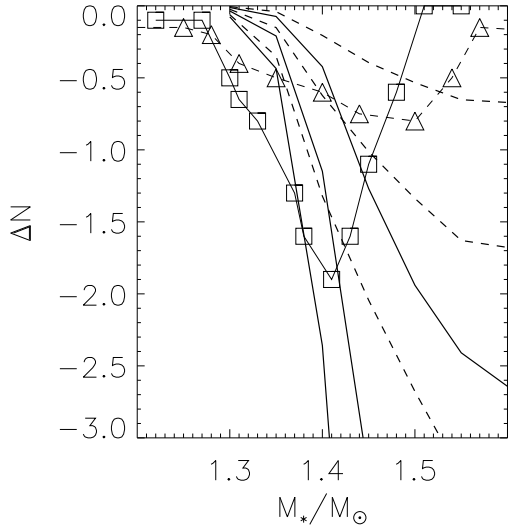,width=8cm}}
\caption{Relative depletion of Li and Be in the Hyades cluster (assumed to be 700Myr old) as a
  function of stellar mass around the Li dip. 
  Note that $\Delta N({\rm Li}) = \log (X_{\rm Li}^{\rm cz}/X_{\rm Li,0})$ 
  and similarly for Be.  The thin lines connecting
  the symbols  show data interpolated from Boesgaard (2005). The solid
  line and squares correspond to the Li data, while the dashed lines
  and triangle correspond to the Be data. The model predictions are
  shown in thick solid and dashed lines for the Li  and Be cases
  respectively, from top to bottom in each case for $a_2 = 0.002$,
  $0.005$ and $0.01$ (the larger $a_2$, the larger the depletion). }
\label{fig:libenodiff}
\end{figure}

Our theoretical results, now expressed as Li and Be depletion fractions, are 
shown in Figure \ref{fig:libenodiff}, assuming values 
of $a_2=0.002$, $0.005$, and $0.01$. They are compared with 
observed Li and Be abundances in the Hyades as reported by Boesgaard (2005).
It is quite clear that contrary to our naive expectation of \S\ref{sec:cart},
the gyroscopic pumping of Li and Be out of the outer convection zone 
does not decrease for the higher-mass stars despite the fact that their
outer convection zone becomes smaller and smaller. This is also clear from
Figure \ref{fig:wrzs}, where $W_{\rm rz}^{\rm spher}$ continues to 
increase with stellar mass instead of being quenched as initially expected. 
In the next sections, we discuss why this is the case,
and how to reconcile the model with observations. 

\subsubsection{Discussion}
\label{sec:lidisc}

The most important conclusion from this analysis is that gyroscopic pumping
alone may be able to explain the cool side of the dip for reasonable values of
the differential rotation rate (one merely needs to adjust $a_2$), 
but always vastly over-estimates the depletion 
rate for stars on the hot side of the dip. 
Since the gyroscopic pumping mechanism is quite generic, and 
arises from simple first-principles of angular momentum conservation
our study then raises the question of how one might {\it suppress} 
the effects of pumping in the hot side of the dip. 
In order to address the problem, it is important first to understand the 
cause of the sharp decrease in the predicted depletion timescale for high-mass
stars in this model. 

\begin{figure}[h]
\centerline{\epsfig{file=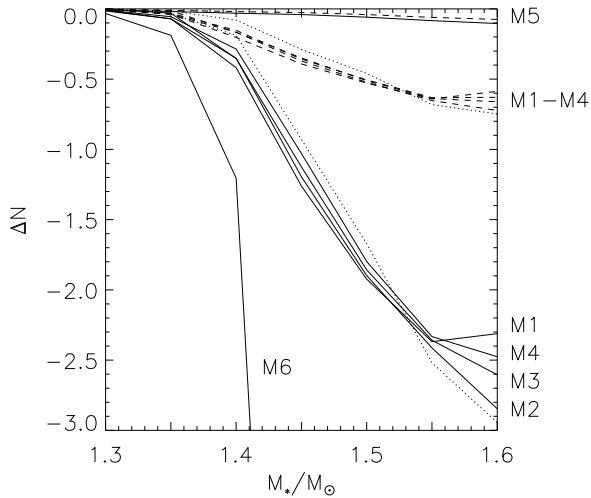,width=8cm}}
\caption{Relative depletion of Li (solid lines)  and Be (dashed lines)
  in the Hyades cluster as a  function of stellar mass in the Li dip,
  as predicted in various artificial test-models. $\Delta N$ is defined as 
  in Fig. \ref{fig:libenodiff}. In all cases, $a_2 =
  0.002$. In model M1, $\Lambda_{\rm out}$ is set to zero for all
  stars. In model M2, $\Omega_\star = 10^{-4}$rad/s for all stars. The
  dotted lines show the variant of model M2 for which $\Omega_\star =
  2 \times 10^{-5}$rad/s and $a_2 = 0.08$, for Li and Be
  respectively. In model M3, $\Lambda_{\rm in}= 10$ for all stars. In
  model M4, $r_{\rm in} = 0.05R_\star$ for all stars. In model M5, the
  mass of the outer convective zone is set $10^{-3} M_\odot$ for all
  stars. In model M6, the contribution of the cylinder (see main text)
  to $M_{\rm Li}$ and $M_{\rm Be}$ is set to 0. All other quantities
  are as in Figure \ref{fig:libenodiff}. Note that in model M6, the two
  lines for Li and Be are on top of each other, as expected from the
  analytical model.}
\label{fig:libestudy}
\end{figure}

First, note that most of the pumping in the stars considered comes from the 
inner core. Indeed, the density in the outer convection zone is so 
low that the mass flux $\bar \rho u_{\rm cz}^{\rm out}$ 
generated in the outer convection zone is negligible compared with the mass
flux generated from the inner convection zone 
(see equation (\ref{eq:wrzconst})). As a result, 
the characteristics of the inner core dominate the {\it forcing} of the flows, 
while the outer convection zone merely plays the role of providing 
a pathway for the flows to return to the interior. 
This is easily verified numerically by setting (artificially) 
$\Lambda_{\rm out} = 0$ to suppress the forcing in the outer convection 
zone. Since turbulent viscosity in the outer
convection zone is nevertheless still present, the return path still exists
and as seen in Figure \ref{fig:libestudy}, the resulting 
depletion rates are hardly changed.

One may then wonder what the main factor controlling the 
variation in the depletion rate as a function of stellar mass actually 
is. A few immediate possibilities come to mind.
The convective velocities in the inner
core of these stars, as well as the size of the inner core (see Table 1), 
both increase with $M_\star$, implying that the total mass flux pumped 
by the inner convection zone is larger for stars on the hot side of the dip. 
In addition, the increase in the rotation rates of the stars as 
$M_\star$ increases implies that the allowed flow speeds through the radiative 
zone are larger on the hot side of the dip (see equation (\ref{eq:esflows})). 
All of these effects explain the trend seen in Figure \ref{fig:wrzs}, which 
clearly shows that the pumped mass flux 
$W_{\rm rz}^{\rm spher}$ increases with $M_\star$ across the dip.

However, this is not sufficient to explain the vast increase in the 
depletion rates observed in Figure \ref{fig:libenodiff} 
as $M_\star$ increases.
In Figure \ref{fig:libestudy} we show different
artificial models to illustrate this statement. 
Model M1 is the aforementioned case where 
$\Lambda_{\rm out}$ is set to zero. Model 
M2 is created holding $\Omega_\star$ constant 
and equal to $10^{-4}$rad/s across all stars in the model, to 
suppress the effect of increased rotation rate across the dip. 
Model M3 is created holding $\Lambda_{\rm in}$ constant 
and equal to 10 across all stars in the model, to suppress the 
effect of increased convective velocities across the dip. Model M4 
is created holding $r_{\rm in}$ constant 
and equal to 0.05$R_\star$ across all stars in the model, to suppress
the effect of increased core size across the dip. In all 
cases, all other quantities
are the same as the original model of Figure \ref{fig:libenodiff}. 
As we can see,
none of these changes affect the predicted depletion rates much.
This incidentally also shows that the exact details of the model 
are not particularly important, and that another, much more fundamental
effect controls the overall depletion rate. 
Finally, we show (as the two dotted lines) 
a variant of model M2 in which $\Omega_\star$ is 
held constant but this time equal to $2\times 10^{-5}$rad/s. This value is 
closer to the typical angular velocity (derived from $v\sin i$) 
of stars for which Li has actually been observed, which is 
significantly smaller than the median rotation rates for stars of the 
same mass (see Boesgaard 1987). Very similar depletion rates 
to those of models M1-M4 can be recovered provided the overall
differential rotation is chosen to be larger ($a_2 = 0.08$). This shows
that there is some degeneracy in the model parameters, which is not 
entirely surprising. 

The dominant effect in the model depletion trend is in fact found 
to be the decrease in the mass of the regions which need to be depleted
in Li or Be as $M_\star$ increases. This mass has two contributions: 
the mass of the outer 
convective zone $M_{\rm out}^{\rm cz}$, plus the mass contained in 
the cylinders $C_{\rm Li}$ and $C_{\rm Be}$ respectively. To 
illustrate the effect in question, we create two additional artificial 
models. In model M5 the mass of the outer convective zone is artificially 
held constant (and equal to $10^{-3} M_\odot$). In model M6 the mass in the 
cylinders is set to 0 (assuming that only the Li and Be fractions 
in the convective zone need to be recirculated). The results are also
shown in Figure \ref{fig:libestudy}. The depletion rates in each case 
are now strikingly different from those of models M1-4. 
In the case of M5, 
the depletion fraction is now much more constant across all stars. 
This strongly suggests that the depletion timescale in 
Figure \ref{fig:libenodiff}  varies with stellar mass
more because the total mass which needs to be depleted 
varies than because the pumped mass flux  $W_{\rm rz}^{\rm spher}$ varies. 
Model M6 illustrates a fundamental property of these 
rapidly rotating Li-dip stars undergoing gyroscopic pumping.
Since the pumped mass flux is independent of depth 
below the convection zone, unless the mass of Be to be depleted
is different from the mass of Li to be depleted, the predicted
depletion timescales will be exactly the same for the two species. This is 
precisely the case illustrated by model M6 which ignores the contribution 
of the cylinders $C_{\rm Li}$ and $C_{\rm Be}$, so the predicted
depletion fractions of Li and Be are exactly the same. 
Hence, the mixed inner cylinders are crucial to the difference
in the depletion timescales of Li and Be. Moreover, their radii and heights
uniquely determine the predicted ratio of the Li and Be 
depletion fraction. As seen in Figure \ref{fig:libenodiff}, the original 
model described in \S\ref{sec:lipred1}
appears to predict the correct ratio, even though 
the absolute depletion fractions are too large for the hot side of the dip. 

The remaining question is why the pumping is not suppressed for 
stars of masses higher than 1.5$M_\odot$, as originally expected from the 
gradual disappearance of the outer convective zone. In fact, it turns
out that $d_{\rm out}$ and $v_{\rm conv}^{\rm out}$ remain significant in these
stars, even though the total mass of the convection zone becomes negligible. 
Since the main role of the outer convection zone is to provide
a pathway to recirculate the small 
mass flux pumped by the inner core, there is no notable suppression 
of the pumping in this mass range contrary to our original idea 
described in \S\ref{sec:cart}. 

\subsection{The role of Li (and Be) diffusion} 

While the model as it stands appears to reproduce the depletion trend on the 
cool side of the dip with reasonable assumptions for the differential 
rotation of the inner core ($a_2 = O(0.01)$), 
it vastly overestimates the depletion fraction
on the hot side of the dip. Our efforts now shift 
to the problem of reconciling the model with observations, or in other words 
on {\it reducing} the depletion rate on the hot side of the dip.

The answer, as it happens, is quite simple, and
lies in the balance between 
advection of Li-rich (and Be-rich) material out of the outer 
convection zone by gyroscopic pumping, 
and diffusion of these elements back into it from the radiative zone below
which, as illustrated in Figure \ref{fig:gyrofinal}, still has 
primordial Li and Be abundances. 
To take this new effect into account, equation (\ref{eq:mlicz1}) 
must be modified as 
\begin{equation}
\dot{M}^{\rm cz}_{\rm Li} =  - \pi r_{\rm in}^2 W^{\rm spher}_{\rm rz}  X^{\rm cz}_{\rm Li} + \pi r_{\rm in}^2 W^{\rm spher}_{\rm rz} X^{\rm cyl}_{\rm Li}  - 4\pi r_{\rm out}^2 \bar \rho (r_{\rm out}) D \nabla X_{\rm Li} \mbox{  ,} \nonumber \\
\label{eq:mlidot2}
\end{equation}
where the new third term is the mass flux of Li diffused back into the 
outer convection zone. The diffusion 
coefficient $D$ is presumably the sum of a microscopic and a 
turbulent contribution from convective overshoot. 
The microscopic contribution\footnote{Note that given all other 
approximations made in this section, the 
effects of radiative levitation and gravitational settling are 
neglected for simplicity.}  is evaluated using the 
prescription of Gough (2007), while the turbulent contribution presumably decays 
exponentially with depth below the convection zone on a typical 
overshoot depth lengthscale, yielding
\begin{equation}
D(r) = D_{\rm micro}(r) + D_{\rm turb}(r) = D_{\rm micro}(r) 
+ v_{\rm conv}^{\rm out}d_{\rm out} \exp\left(\frac{r-r_{\rm out}}{\Delta_{\rm out}}\right) 
\mbox{   for  } r < r_{\rm out} \, .
\end{equation}
The equation for the evolution of $M_{\rm Li}^{\rm cyl}$ remains unchanged 
(neglecting the diffusion of Li through the sides of the cylinder for 
simplicity). 

Unfortunately, the addition of a diffusive component to the Li flux 
now prevents us from deriving 
simple analytical solutions of these equations analogous to the ones
found in \S\ref{sec:lipred1}. Instead, solutions 
can only be obtained numerically by integrating the 
partial differential equation (\ref{eq:mlidot2}) with time, and following the
evolution of the spatially varying Li profile. 
Given the other approximations made throughout this work (e.g. assuming that 
the rotation rate of the star, the background stellar structure and 
the convective forcing are all constant with time), and since our aim 
is merely a preliminary ``proof-of-concept'', it seems futile
to try to obtain a precise numerical solution of (\ref{eq:mlidot2}). 

Instead, in this first paper we proceed by approximating the gradient term to cast 
(\ref{eq:mlidot2}) in the form of an ordinary differential equation similar to 
(\ref{eq:mlidot1}). This can be done simply be writing 
\begin{equation}
\nabla X_{\rm Li} = \frac{X_{\rm Li}^{\rm cz} - X_{\rm Li,0}}{d_{\rm diff}} \mbox{   ,}
\end{equation}
where we have assumed that there is a ``reservoir'' of material with primordial
Li abundance in the radiative zone in the form of a spherical shell of width 
$d_{\rm diff}$ adjacent to $r_{\rm out}$, and that
Li has to diffuse across that reservoir to be mixed 
into the convection zone. For this reservoir to contain enough Li to 
replenish the convection zone it must have roughly the same
mass, so we set $d_{\rm diff} = \beta d_{\rm out}$ 
where $\beta$ is a free parameter of the problem of order unity. 
Finally, the diffusion coefficient must also be approximated
by a constant for (\ref{eq:mlidot2}) to become a true ODE. 
Since the turbulent transport is 
weakest furthest from the radiative--convective interface, 
this is the ``bottleneck'' region for the diffusion of Li
back into the convection zone. Hence we take 
\begin{equation}
D = D_{\rm micro}(r_{\rm out} - d_{\rm diff}) + v_{\rm conv}^{\rm out}d_{\rm out} \exp\left(-\frac{d_{\rm diff}}{\Delta_{\rm out}}\right) \mbox{   .}
\end{equation}

As before, we divide (\ref{eq:mlidot2}) by $M_{\rm cz}^{\rm out}$ to get
\begin{eqnarray}
&& \dot{X}^{\rm cz}_{\rm Li} \simeq  -\frac{ X^{\rm cz}_{\rm Li}}{\tau^{\rm cz}_{\rm pump}} + \frac{ X^{\rm cyl}_{\rm Li}}{\tau^{\rm cz}_{\rm pump}} - \frac{ X_{\rm Li}^{\rm cz} - X_{\rm Li}(0) }{\tau_{\rm diff}} \mbox{  ,} \nonumber \\
\label{eq:xlidot2}
&& \dot{X}^{\rm cyl}_{\rm Li} \simeq  \frac{ X^{\rm cz}_{\rm Li}}{\tau^{\rm cyl}_{\rm Li,pump}} - 2\frac{ X^{\rm cyl}_{\rm Li}}{\tau^{\rm cyl}_{\rm Li,pump}}\mbox{  ,}
\end{eqnarray}
where 
\begin{equation}
 \tau_{\rm diff} = \frac{D}{d_{\rm diff} d_{\rm out}} \frac{4 \pi r_{\rm out}^2 d_{\rm out} \bar \rho(r_{\rm out}) }{M_{\rm cz}^{\rm out}}  \simeq \frac{D}{\beta d_{\rm out}^2}\mbox{  .}
\end{equation}
As shown in Appendix B, one can integrate these equations 
analytically fairly straightforwardly. We find, as expected on physical grounds,
that when the Li diffusion timescale into the convection 
zone becomes shorter than the advection timescale out of the convection zone, 
the surface Li abundance remains close to the primordial value. 

\begin{figure}[h]
\centerline{\epsfig{file=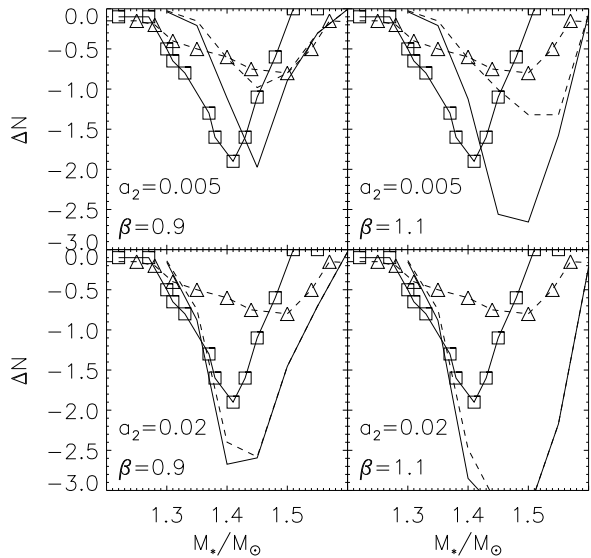,width=8cm}}
\caption{Relative depletion of Li (solid lines) 
and Be (dashed lines) in the Hyades cluster as a 
function of stellar mass in the Li dip, as predicted for various 
values of the model parameters $a_2$ (related to the stellar 
differential rotation) and $\beta$ (related to the diffusion 
of chemical species across the base of the convection zone). $\Delta N$ is defined as in Fig. \ref{fig:libenodiff}. See main text for detail. }
\label{fig:libewithdiff}
\end{figure}

The resulting depletion fractions calculated using this method, 
for Li and Be, are shown in Fig. \ref{fig:libewithdiff} for a 
simple grid of parameters 
($a_2 = 0.005$, $a_2 = 0.02$, $\beta = 0.9$ and $\beta = 1.1$), with the
stellar models otherwise exactly the same as the ones used to create 
Figure \ref{fig:libenodiff}. 
Note how, for all chosen parameter values, the Li and Be profiles now 
exhibit a clear dip centered roughly around the position of the observed Li-dip. 
On the cool side the effects of diffusion are negligible and the predicted 
depletion rates are very similar to the ones obtained in the absence of 
diffusion (see Figure \ref{fig:libenodiff}). 
On the hot side, diffusion is important and continuously replenishes the outer 
convection zone with Li-rich and Be-rich material. The transition between the cool 
and hot sides, in the model, occurs when the diffusion timescale 
of Li back into the convection zone, namely $\tau_{\rm diff}$,
becomes comparable with the advection timescale of Li-rich material out 
of it (namely $\tau^{\rm cz}_{\rm Li,pump}$), 
and depends both on the diffusion rate (as controlled by $\beta$ for 
example) and on the pumping rate (as controlled by $a_2$). 

Note that for the larger value of $a_2$, the ratio of the 
Li depletion fraction to the Be depletion fraction within the dip 
remains close to 1, contrary to observations.
Instead, the ratio of the 
predicted depletion fractions is closer to observations for lower 
values of $a_2$. The overall ``best fit'' is found for $a_2 = 0.005$ 
and $\beta = 0.9$, which are not unreasonable parameter values. However,
with our very crude model of the diffusion, we are not able to fit 
the exact position and amplitude of {\it both} Li and Be dips simultaneously.
Since we were only aiming for a proof-of-concept, we view the results
obtained as quite satisfactory, although a more careful model will be required 
in the future should one wish to explain the structure of both dips in more detail. 

\subsection{Summary}

To summarize this section, we have found that the combination of
gyroscopic pumping {\it and} turbulent diffusion of chemical species 
by overshooting motions ubiquitously predicts the presence of a dip in 
both Li and Be surface abundances for young MS stars in the mass 
range $1.3-1.5M_\odot$ (Li-dip stars). 
Depletion profiles close to the observed ones
can be reproduced for reasonable values of model parameters.
The increase in the depletion fraction on the 
cool side of the dip is explained by gyroscopic pumping and the 
progressively smaller amount of material which needs to be recirculated, 
while the decrease in the depletion fraction on the hot side of the dip 
is explained by the effect of diffusion on replenishing the outer convection 
zone with Li and Be. 

\section{Discussion and prospects}
\label{sec:ccl}

In this paper we performed an exhaustive study of a non-local
source of rotational mixing called ``gyroscopic pumping'', which was originally
studied in the context of the Earth's atmosphere by Haynes et al. (1991) 
and later discussed in the case of the Sun by Gough \& McIntyre (1998),
McIntyre (2007), and GAA09. 

In this mechanism, large-scale meridional motions are driven by 
angular-momentum conservation whenever fluid undergoes forces in the 
azimuthal direction (see \S\ref{sec:intromix} and in particular 
Figure \ref{fig:gyro} for detail). This is exactly the case in stellar 
convective zones, where rotationally influenced convection gives
rise to the so-called $\Lambda-$effect (cf. R\"udiger 1989; see also 
Garaud et al. 2010), and typically leads to the 
azimuthal acceleration of equatorial regions and deceleration of the poles.
The resulting large-scale meridional circulation has fluid flowing 
outward from the rotation axis near the equator, and toward the rotation 
axis near the pole. 
Note that this mechanism is quite generic, since it is simply based on 
angular momentum conservation. However, by contrast with other 
well-studied significant sources of rotational mixing, 
it does not rely on stellar spin-down to be effective -- 
it is an inherently quasi-steady mechanism. 

As first shown by GAA09 and studied more extensively here, 
whether the fluid gyroscopically pumped in the convective zone 
ends up mixing the nearby radiative zone or not depends on many factors. 
The case of stars with a single convection zone, in the absence of any 
other mechanism, was first studied by GAA09. In this case the overall 
amplitude of the flows penetrating into the radiative zone is 
limited to slow Ekman flows, which are unlikely 
to play any significant role in mixing the stellar interior. However, this
conclusion could be different if a large-scale magnetic field is present 
(Gough \& McIntyre 1998), as it is for instance thought to be the case in the Sun.
We defer the magnetic case to a subsequent paper. 

Here we focused on stars with two convection zones, 
in the absence of magnetic fields (we accept that this 
approximation is probably over-simplistic). 
We found that gyroscopic pumping
provides a significant source of non-local mixing in the radiative zones of
these stars, more precisely along the rotation axis, within the 
cylinder tangent to the inner core (see Figure \ref{fig:gyrofinal}). 
This mixing readily explains, for example,
the cool side of the well-known Li-dip and, when moderated by the effects of 
diffusion, provides predictions for the Li- and Be-depletion fractions which 
give a surprisingly good fit to the data on both sides of the dip 
(see Figure \ref{fig:libewithdiff}). It is important to note that contrary
to previous models, turbulent mixing here is needed not to explain Li 
destruction but to explain the replenishment of the surface layers in Li. 
Of course, the various other effects described in \S\ref{sec:liintro}, 
which were not taken into account here, could also affect 
the Li and Be abundances: mixing by gravity waves, large-scale rotational
mixing induced by stellar spin-down, radiative acceleration, etc. A more
sophisticated study of the Li-dip in the light of gyroscopic pumping, 
taking these effects into account as well as stellar evolution, 
is deferred to a subsequent publication. Furthermore, it is interesting
to note that the gyroscopic pumping mechanism probably plays a role in the
global redistribution of angular momentum within the star. Whether 
this is related to the dichotomy between slow rotators and fast rotators
on either side of the dip is an interesting question, which we hope 
to address in the future.

Gyroscopic pumping, as studied here, also opens up other interesting 
observational prospects beyond Li-dip stars. For example, another important
evolutionary phase when      stars have two convective zones is the core helium 
burning phase on the red giant branch. 
Our study shows that this pumping mechanism could 
provide an important connection between the convective envelope and 
the core, and perhaps help reconciliate some of the long-standing discrepancies in 
surface abundances of CNO elements between models and observations. 
It is also interesting to note that the pumping itself has a tendency
to drive a strong azimuthal shear in the system, close to the tangent 
cylinder (see Figure \ref{fig:sample}). This shear could perhaps become
unstable for very rapidly rotating stars, leading to much more violent
mixing in the radiative zone. Alternatively, this shear layer could also
become the seat of an $\Omega-$effect and generate strong toroidal fields
around the tangent cylinder. It is interesting to speculate on what 
kind of field such a flow structure would sustain, and what observable
feature it may lead to -- although a
full answer to that question will require a study of 
gyroscopic pumping in the presence of magnetic fields. 
\nocite{z92,wa97,tz97,tc03,tc98,sz92,sal93a,sal93b,sal10,sal90,sal98,r89,rm93,r07,p97,mci07,m86,k90,hal91,g07,gmi98,gls91,gg08,2001PhDT,gp08,gaa09,gal10,dp97,cc07,jcdal96,bk02,bt86,b05,b87,bal07,bal96,atal09}

\section*{Acknowledgements}

This work originated from a proposed project at the Woods
Hole GFD Summer School in 2009, which was unfortunately not
selected by any of the summer fellows. Nevertheless, P. Garaud
thanks the NSF and the ONR for supporting
this excellent program. P. Garaud was supported by an NSF CAREER award. 
The numerical simulations were performed on the Pleiades cluster at UCSC, 
purchased using an NSF-MRI grant. We thank N. Brummell, P. Charbonneau 
and G. Michaud for fruitful discussions.

\section*{Appendix A: Analytical derivation of $W_{\rm rz}$}

In this appendix, we derive the analytical expression for the 
solution of the Cartesian model described in \S\ref{sec:analytical}, 
focusing in particular on the derivation of the 
mass flux into the radiative zone, $W_{\rm rz}$. 
Solutions to the set of equations (\ref{eq:maineqs}) are first found in each of the three
regions separately, then matched to one another at the interfaces (at $z_{\rm in}$ and $z_{\rm out}$)  
and to boundary conditions (at $z=0$ and $z=1$). 

\paragraph{Solution in the outer convection zone.}

In the outer convection zone, the viscous stresses are
negligible compared with the linear drag term and $N(z) = 0$. 
As a result, the equations reduce to 
\begin{eqnarray}
\hat T_{zz} - k^2 \hat T = 0 \mbox{   ,  }  \nonumber \\
 - 2\hat v = - \Lambda_{\rm out} ( \hat u - \hat u^{\rm out}_{\rm cz})  \mbox{   ,  } \nonumber \\
 2\hat u = - ik e^{z/D_\rho} \hat p - \Lambda_{\rm out} \hat v  \mbox{   ,  } \nonumber \\
 0 = - \hat p_z e^{z/D_\rho} + e^{z/D_T} \hat T  - \Lambda_{\rm out} \hat w \mbox{   ,  } \nonumber \\
e^{-z/D_\rho}ik \hat v + (e^{-z/D_\rho} \hat w)_z =0  \mbox{   . }
\end{eqnarray}
The temperature equation is easily solved as 
\begin{equation}
\hat T(z) = a e^{kz} + b e^{-kz}  \mbox{   .  }
\end{equation}
The remaining system of four equations can by transformed into 
a linear system of ODEs with {\it constant} coefficients using the following
new mass flux variables: $ U = e^{-z/D_\rho} \hat u$ (and similarly for $u_{\rm cz}$), 
$ V = e^{-z/D_\rho} \hat v$, and $ W = e^{-z/D_\rho} \hat w$. In that case, 
\begin{eqnarray}
&&-2  V =  - \Lambda_{\rm out} ( U -  U^{\rm out}_{\rm cz}) \mbox{   ,   }  \nonumber \\
&&2 U = - ik \hat p  - \Lambda_{\rm out}  V \mbox{   ,   }  \nonumber \\
&&0 = - \hat p_z - \Lambda_{\rm out} W +  T e^{-z/L} \mbox{   ,   }  \nonumber \\
&& ik  V + W_z = 0 \mbox{   ,   }
\end{eqnarray}
where $L^{-1} = D_{\rho}^{-1} - D_T^{-1}$. 

Eliminating each variable in turn we can reduce the system to 
a second-order, forced linear ordinary differential equation 
for $W$ (for example): 
\begin{equation}
W_{zz} = W   \frac{k^2\Lambda_{\rm out}^2}{4+\Lambda_{\rm out}^2} + \frac{2ik\Lambda_{\rm out}}{4+\Lambda_{\rm out}^2} S_{\rm cz}^{\rm out}  - e^{-z/L} \frac{k^2\Lambda_{\rm out}}{4+\Lambda_{\rm out}^2} \hat T  \mbox{   ,} 
\end{equation}
where
\begin{equation}
S_{\rm cz}^{\rm out}(z) = \frac{\dd U^{\rm out}_{\rm cz}(z)}{\dd z} =\frac{\dd }{\dd z} \left(e^{-z/D_\rho} \hat u^{\rm out}_{\rm cz}(z)\right)  \mbox{   .}
\end{equation} 
The solution of this equation is the sum of the solution of the homogeneous problem plus a particular solution. 
If we assume for simplicity that $S_{\rm cz}^{\rm out}$ is constant in the
 outer convection zone\footnote{Of course, this is not the case for the
 example chosen here. However, the small difference in the solution caused by
 the non-constant $S_{\rm cz}^{\rm out}$ is not worth the increased
 complication
  in the algebra since we are looking here only at obtaining a physical intuition of the solution.} , the particular 
solution is easily expressed as: 
\begin{equation}
W(z) = A e^{z/\delta_{\rm out}} + B e^{-z/\delta_{\rm out}} -\frac{2iS_{\rm cz}^{\rm out} }{k\Lambda_{\rm out}} - \frac{a}{\Lambda_{\rm out}} \frac{ e^{(k-L^{-1})z}}{\delta_{\rm out}^2 (k-L^{-1})^2 - 1}  - \frac{b}{\Lambda_{\rm out}} \frac{e^{(-k-L^{-1})z}}{\delta_{\rm out}^2 (k+L^{-1})^2 - 1} \mbox{   ,}
\end{equation}
where we have defined the new lengthscale
\begin{equation}
\delta_{\rm out} = \frac{\sqrt{4+\Lambda_{\rm out}^2}}{k\Lambda_{\rm out}} \mbox{   .}
\end{equation}
Using the solutions obtained for $W$ and $T$ we deduce the solution for the pressure perturbation: 
\begin{eqnarray}
\hat p(z)&=& \frac{2iU^{\rm out}_{\rm cz} }{k} - \Lambda_{\rm out} \delta_{\rm out} \left( A e^{z/\delta_{\rm out}} - B e^{-z/\delta_{\rm out}}\right) \nonumber \\
 &+&  a \frac{(k-L^{-1}) e^{(k-L^{-1})z}}{(k-L^{-1})^2 - \delta_{\rm out}^{-2}}  -  b \frac{(k+L^{-1})e^{(-k-L^{-1})z}}{(k+L^{-1})^2 - \delta_{\rm out}^{-2}} \mbox{   .}
\end{eqnarray}

\paragraph{Solution in the inner convection zone.}

By analogy, the solutions in the inner convection zone are given by
\begin{eqnarray}
\label{eq:inner_conv_sol}
&& \hat T(z) = c e^{kz} + d e^{-kz} \nonumber \\
&& W(z) = C e^{z/\delta_{\rm in}} + D e^{-z/\delta_{\rm in}} -\frac{2iS_{\rm cz}^{\rm in} }{k\Lambda_{\rm in}} - \frac{ c}{\Lambda_{\rm in}} \frac{ e^{(k-L^{-1})z}}{\delta_{\rm in}^2 (k-L^{-1})^2 - 1}  - \frac{ d}{\Lambda_{\rm in}} \frac{e^{(-k-L^{-1})z}}{\delta_{\rm in}^2 (k+L^{-1})^2 - 1}  \mbox{   ,}
\nonumber \\
&& \hat p(z) = \frac{2iU^{\rm in}_{\rm cz}}{k} - \Lambda_{\rm in} \delta_{\rm in} \left( C e^{z/\delta_{\rm in}} - D e^{-z/\delta_{\rm in}}\right) \nonumber \\ && \quad \quad \quad + c \frac{(k-L^{-1}) e^{(k-L^{-1})z}}{(k-L^{-1})^2 - \delta_{\rm in}^{-2}}  -  d \frac{(k+L^{-1})e^{(-k-L^{-1})z}}{(k+L^{-1})^2 - \delta_{\rm in}^{-2}}\mbox{   ,}
\end{eqnarray}
where $\delta_{\rm in}$ and $S_{\rm cz}^{\rm in}$ are defined by analogy with
$\delta_{\rm out}$ and $S_{\rm cz}^{\rm out}$ and
where four new integration constants ($c$, $d$, $C$ and $D$) 
have been introduced. 

\paragraph{Solution in the radiative zone.}

In the radiative zone, the governing equations (\ref{eq:maineqs}) reduce to 
\begin{eqnarray}
&&-2 \hat v =  E_\nu \left(\hat u_{zz} - k^2 \hat u \right) \mbox{   ,   }  \nonumber \\
&&2 \hat u = - ik \hat p e^{z/D_\rho} + E_\nu \left( \hat v_{zz} - k^2 \hat v \right) \mbox{   ,   }  \nonumber \\
&&0 = - \hat p_z e^{z/D_\rho} + \hat T e^{z/D_T} + E_\nu  \left(\hat w_{zz} - k^2 \hat w \right) \mbox{   ,   }  \nonumber \\
&& \frac{N_{\rm rz}^2}{\Omega_\star^2}  e^{-z/D_T} \hat w = E_\kappa  \left( \hat T_{zz}- k^2 \hat T \right) \mbox{   ,   } \nonumber \\
&& ik  \hat v e^{-z/D_\rho} + (e^{-z/D_\rho}\hat w)_z = 0 \mbox{   .   }
\label{eq:maineqsrad}
\end{eqnarray}
In the limit of very small Ekman number, and for $\sigma <1$,  the system is approximated by 
\begin{eqnarray}
 - 2\hat v =  O(E_\nu)  \mbox{  , }  \nonumber \\
 2\hat u = - ike^{z/D_\rho} \hat p+ O(E_\nu)   \mbox{  , } \nonumber \\
e^{z/D_\rho} \hat p_z  = e^{z/D_T}\hat T + O(E_\nu)   \mbox{  , }  \nonumber \\
e^{-z/D_T} \frac{N_{\rm rz}^2}{\Omega_\star^2  }\hat w = E_\kappa (\hat T_{zz} - k^2 \hat T)   \mbox{  , } \nonumber \\
(e^{-z/D_\rho}\hat w)_z = O(E_\nu)   \mbox{  , } 
\end{eqnarray}
which successively implies that 
\begin{eqnarray}
 e^{-z/D_\rho}\hat w = W_{\rm rz} + O(E_\nu)   \mbox{  , } \nonumber \\
\hat T = K_1 e^{kz} + K_2 e^{-kz} + \frac{N_{\rm rz}^2}{\Omega_\star^2  }  \frac{W_{\rm rz}}{E_\kappa} \frac{e^{z/L} }{L^{-2}-k^2 } + O(E_\nu)   \mbox{  , }  \nonumber \\
\hat p = p_{\rm rz} +  \frac{K_1 e^{z (k-L^{-1})} }{k-L^{-1}} -  \frac{K_2   e^{z(-k-L^{-1})} }{k+L^{-1}}  + \frac{N_{\rm rz}^2}{\Omega_\star^2 }\frac{W_{\rm rz}}{ E_\kappa}  \frac{z}{ L^{-2}- k^2} + O(E_\nu)   \mbox{  , } 
\end{eqnarray}
and where $W_{\rm rz}$, $K_1$, $K_2$ and $p_{\rm rz}$ are integration constants. 

We see that the condition where $\bar \rho \hat w$ is constant arises simply from mass conservation 
and geostrophy. 
Assuming that $W_{\rm rz}$ is known, the pumped flow induces local temperature perturbations 
as a result of the thermal energy equation: this can easily be seen in the solution for $\hat T$, 
which contains two parts: the particular solution of the Poisson 
problem which includes the source term arising from the advection of the background temperature 
by the meridional flows, plus the general solution of $\nabla^2 \hat T = 0$. The constants 
$K_1$ and $K_2$ are simply there to match the temperature profile in the 
radiative zone to that of the convection zones. The expression for the pressure perturbation 
can be interpreted in a similar way. 

\paragraph{Matching the solutions and radial velocity in the radiative 
zone.}

We have found twelve integration constants, which can be uniquely determined
by applying four boundary conditions (impermeability 
at $z=0$ and  $z=1$, and $\hat T= 0$ at the top and the bottom) 
and eight matching conditions (continuity of 
$\hat w$ -- alternatively of $W$ -- $\hat p$, $\hat T$ and $d\hat T/dz$ at each of the two interfaces).

The boundary conditions at $z=0$ and  $z=1$ imply: 
\begin{eqnarray}
A e^{1/\delta_{\rm out}} + B  e^{-1/\delta_{\rm out}} = \frac{2i S_{\rm cz}^{\rm out}}{k \Lambda_{\rm out}}  + \frac{a}{\Lambda_{\rm out}} \frac{ e^{(k-L^{-1})}}{\delta_{\rm out}^2 (k-L^{-1})^2 - 1}  + \frac{b}{\Lambda_{\rm out}} \frac{e^{(-k-L^{-1})}}{\delta_{\rm out}^2 (k+L^{-1})^2 - 1}    \mbox{  , } \nonumber \\
C + D  = \frac{2i S_{\rm cz}^{\rm in}}{k \Lambda_{\rm in}} + \frac{c}{\Lambda_{\rm in}} \frac{ 1}{\delta_{\rm in}^2 (k-L^{-1})^2 - 1}  + \frac{ d}{\Lambda_{\rm in}} \frac{1}{\delta_{\rm in}^2 (k+L^{-1})^2 - 1}   \mbox{  , } \nonumber \\
a e^{k} + b  e^{-k} = 0   \mbox{  , } \nonumber \\
c + d = 0   \mbox{  . } 
\end{eqnarray}

Continuity of vertical velocity, pressure, temperature and derivative of temperature at the $z=z_{\rm out}$ interface implies
\begin{eqnarray}
&& A e^{z_{\rm out}/\delta_{\rm out}} + B  e^{-z_{\rm out}/\delta_{\rm out}} - \frac{2iS_{\rm cz}^{\rm out}}{k \Lambda_{\rm out}} - \frac{a}{\Lambda_{\rm out}} \frac{ e^{(k-L^{-1})z_{\rm out}}}{\delta_{\rm out}^2 (k-L^{-1})^2 - 1}  - \frac{b}{\Lambda_{\rm out}} \frac{e^{(-k-L^{-1})z_{\rm out}}}{\delta_{\rm out}^2 (k+L^{-1})^2 - 1} =  W_{\rm rz}   \mbox{  , } \nonumber \\
&& \frac{2 i U^{\rm out}_{\rm cz}}{k}  -\delta_{\rm out} \Lambda_{\rm out} \left[ A e^{z_{\rm out}/\delta_{\rm out}} - B e^{-z_{\rm out}/\delta_{\rm out}} \right] + a \frac{(k-L^{-1}) e^{(k-L^{-1})z_{\rm out}}}{(k-L^{-1})^2 - \delta_{\rm out}^{-2}}  -  b \frac{(k+L^{-1})e^{(-k-L^{-1})z_{\rm out}}}{(k+L^{-1})^2 - \delta_{\rm out}^{-2}} \nonumber \\  && \quad \quad \quad\quad =  p_{\rm rz} + \frac{K_1 e^{z_{\rm out} (k-L^{-1})} }{k-L^{-1}}-  \frac{K_2  e^{z_{\rm out}(-k-L^{-1})} }{k+L^{-1}}  +  \frac{N_{\rm rz}^2}{\Omega_\star^2 }\frac{W_{\rm rz}}{ E_\kappa}  \frac{z_{\rm out}}{L^{-2}- k^2}    \mbox{  , }     \nonumber \\
&& a e^{ kz_{\rm out}} + b e^{-k z_{\rm out}} = K_1 e^{kz_{\rm out}} + K_2 e^{-kz_{\rm out}}  + \frac{N_{\rm rz}^2}{\Omega_\star^2  } \frac{W_{\rm rz}}{ E_\kappa} \frac{e^{z_{\rm out}/L} }{ L^{-2}-k^2 }  \mbox{  , }  \nonumber \\
&& k a e^{k z_{\rm out}} -k b e^{-k z_{\rm out}} = k K_1 e^{kz_{\rm out}} -k K_2 e^{-kz_{\rm out}} + \frac{N_{\rm rz}^2}{\Omega_\star^2   } \frac{W_{\rm rz}}{LE_\kappa} \frac{e^{z_{\rm out}/L} }{L^{-2}-k^2 }   \mbox{  . }  
\end{eqnarray}

Similar matching conditions at the lower interface ($z = z_{\rm in}$) imply 
\begin{eqnarray}
&& C e^{z_{\rm in}/\delta_{\rm in}} + D e^{-z_{\rm in}/\delta_{\rm in}} - \frac{2iS_{\rm in}}{k \Lambda_{\rm in}} - \frac{c}{\Lambda_{\rm in}} \frac{ e^{(k-L^{-1})z_{\rm in}}}{\delta_{\rm in}^2 (k-L^{-1})^2 - 1}  - \frac{d}{\Lambda_{\rm in}} \frac{e^{(-k-L^{-1})z_{\rm in}}}{\delta_{\rm in}^2 (k+L^{-1})^2 - 1}  =  W_{\rm rz}    \mbox{  , } \nonumber \\
&& \frac{2 iU^{\rm in}_{\rm cz}}{k}  -\delta_{\rm in} \Lambda_{\rm in} \left[ C e^{z_{\rm in}/\delta_{\rm in}} - D e^{-z_{\rm in}/\delta_{\rm in}} \right] + c \frac{(k-L^{-1}) e^{(k-L^{-1})z_{\rm in}}}{(k-L^{-1})^2 - \delta_{\rm in}^{-2}}  -  d \frac{(k+L^{-1})e^{(-k-L^{-1})z_{\rm in}}}{(k+L^{-1})^2 - \delta_{\rm in}^{-2}}  \nonumber \\ && \quad  \quad \quad \quad  = p_{\rm rz} +  \frac{K_1  e^{z_{\rm im} (k-L^{-1})} }{k-L^{-1}} -  \frac{K_2 e^{z_{\rm in}(-k-L^{-1})} }{k+L^{-1}}  +  \frac{N_{\rm rz}^2}{\Omega_\star^2 }\frac{W_{\rm rz}}{ E_\kappa}  \frac{z_{\rm in}}{L^{-2}- k^2}    \mbox{  , } \nonumber \\
&& c e^{k z_{\rm in}} + d e^{-k z_{\rm in}} = K_1 e^{kz_{\rm in}} + K_2 e^{-kz_{\rm in}} + \frac{N_{\rm rz}^2}{\Omega_\star^2   } \frac{W_{\rm rz}}{E_\kappa} \frac{e^{z_{\rm in}/L} }{L^{-2}-k^2 }   \mbox{  , }    \nonumber \\
&&  kc e^{k z_{\rm in}} - kd e^{-k z_{\rm in}} = kK_1 e^{kz_{\rm in}} - kK_2 e^{-kz_{\rm in}} +  \frac{N_{\rm rz}^2}{\Omega_\star^2   } \frac{W_{\rm rz}}{L E_\kappa} \frac{e^{z_{\rm in}/L} }{L^{-2}-k^2 }    \mbox{  .} 
\end{eqnarray}
This system of twelve equations can in principle be solved analytically 
exactly, but the solutions are horrendously complicated and without much 
interest. To get a better physical intuition of the solutions, we restrict 
our study to stars where the convection zone depths 
$d_{\rm in}$ and $d_{\rm out}$ are small compared with $k$, $\delta_{\rm in}$ and $\delta_{\rm out}$. 
Since $k=2$, and $\delta_{\rm in}$ and  $\delta_{\rm out}$ are always 
greater than one for all possible value of $\Lambda_{\rm in}$ and $\Lambda_{\rm out}$, 
it is sufficient to require that $d_{\rm in}$ 
and $d_{\rm out}$ be much smaller than one. Note that the 
$d_{\rm in} \ll 1$ and 
$d_{\rm out} \ll 1$ approximations are acceptable for all Li-dip stars, 
which have thin outer convective regions and 
small convective cores.
We also note that $L$ is typically small compared with $\delta$ or $1/k$, 
although could be of the same order of magnitude as $d_{\rm in}$ and $d_{\rm out}$.

We begin by solving for the constants $a$, $b$, $c$, $d$, $K_1$ and $K_2$ in terms of $W_{\rm rz}$. This yields, in the limit of thin convection zones: 
\begin{eqnarray}
&& a  = K_1  + \frac{N_{\rm rz}^2}{2\Omega_\star^2  } \frac{W_{\rm rz} L^2 }{E_\kappa} e^{z_{\rm out}/L} e^{ -kz_{\rm out}}  \left( 1 + \frac{1}{kL} \right)   \mbox{  , } \nonumber \\
&& b =  K_2  + \frac{N_{\rm rz}^2}{2\Omega_\star^2   } \frac{W_{\rm rz}L^2 }{E_\kappa}e^{z_{\rm out}/L} e^{ kz_{\rm out}}  \left( 1 - \frac{1}{kL} \right)   \mbox{  , }  \nonumber \\
&& c  = K_1 + \frac{N_{\rm rz}^2}{2\Omega_\star^2   } \frac{W_{\rm rz}L^2}{E_\kappa} e^{z_{\rm in}/L}  e^{-k z_{\rm in}}  \left( 1 + \frac{1}{kL} \right)    \mbox{  , }  \nonumber \\
&&  d  =  K_2  + \frac{N_{\rm rz}^2}{2\Omega_\star^2   }  \frac{W_{\rm rz}L^2}{E_\kappa}  e^{z_{\rm in}/L} e^{k z_{\rm in}}  \left( 1 - \frac{1}{kL} \right)    \mbox{  , } 
\end{eqnarray}
and then
\begin{eqnarray}
K_1 &=& \frac{N_{\rm rz}^2}{\Omega_\star^2  } \frac{W_{\rm rz}}{E_\kappa} \frac{L^2}{1-e^{2k}} \left[ e^{z_{\rm out}/L} e^k \left( 1+ \frac{d_{\rm out}}{L} \right) - e^{z_{\rm in}/L} \left(1 - \frac{d_{\rm in}}{L} \right) \right]   \mbox{  , } \nonumber \\
K_2 &=&  \frac{N_{\rm rz}^2}{\Omega_\star^2  } \frac{W_{\rm rz}}{E_\kappa}  \frac{L^2}{1-e^{-2k}} \left[ e^{z_{\rm out}/L}  e^{-k}\left( 1 + \frac{d_{\rm out}}{L} \right) - e^{z_{\rm in}/L}\left(1 - \frac{d_{\rm in}}{L} \right) \right]    \mbox{  . } 
\end{eqnarray}

Next, we solve for $A$, $B$, $C$ and $D$ in terms of $W_{\rm rz}$:
\begin{eqnarray}
&& Ae^{1/\delta_{\rm out} } = \frac{iS_{\rm out}}{k \Lambda_{\rm out} }   +  e^{-z_{\rm out}/L} \frac{ ae^{k}}{\Lambda_{\rm out} }\frac{ k L^2}{\delta_{\rm out}   }\left(1- \frac{2L}{d_{\rm out}} \right) -  \frac{W_{\rm rz}}{2} \frac{\delta_{\rm out} }{d_{\rm out}}  +  \frac{2a e^k kL^3}{\Lambda_{\rm out}  \delta_{\rm out} d_{\rm out}} e^{-1/L}    \mbox{  , }  \nonumber \\
&& B e^{-1/\delta_{\rm out}}  = \frac{iS_{\rm out}}{k \Lambda_{\rm out}}   -  e^{-z_{\rm out}/L} \frac{ae^k}{\Lambda_{\rm out}}\frac{ k L^2}{\delta_{\rm out}  }\left(1- \frac{2L}{d_{\rm out}} \right) +  \frac{W_{\rm rz}}{2} \frac{\delta_{\rm out}}{d_{\rm out}} - \frac{2a e^k kL^3}{\Lambda_{\rm out} \delta_{\rm out} d_{\rm out}} e^{-1/L}   \mbox{  , } \nonumber \\
&& C =  \frac{i S_{\rm in}}{k \Lambda_{\rm in}}  + e^{-z_{\rm in}/L} \frac{c}{\Lambda_{\rm in}} \frac{kL^2}{\delta_{\rm in} }\left( 1 + \frac{2L}{d_{\rm in}}\right)  +  \frac{W_{\rm rz}}{2} \frac{\delta_{\rm in}}{z_{\rm in}} - \frac{2ckL^3}{z_{\rm in}\Lambda_{\rm in}\delta_{\rm in}}    \mbox{  , }  \nonumber \\
&& D  =  \frac{i S_{\rm in}}{k \Lambda_{\rm in}}  - e^{-z_{\rm in}/L} \frac{c}{\Lambda_{\rm in}} \frac{kL^2}{\delta_{\rm in} }\left( 1 + \frac{2L}{d_{\rm in}}\right)  -  \frac{W_{\rm rz}}{2} \frac{\delta_{\rm in}}{z_{\rm in}} + \frac{2ckL^3}{z_{\rm in}\Lambda_{\rm in}\delta_{\rm in}}   \mbox{  , } 
\end{eqnarray} 
where $a$ and $c$ are explicitly written above.

These expressions can finally be substituted into the pressure continuity equations at both interfaces, yielding two equations for $p_{\rm rz}$ and $W_{\rm rz}$. Eliminating $p_{\rm rz}$ yields (\ref{eq:wrzconst}).  A generalization of this derivation for more complex density, temperature and buoyancy frequency profiles can also be performed analytically although in practice is not particularly useful. For real stellar models, it is much easier to solve (\ref{eq:maineqs}) numerically, although the overall scaling of the solutions with parameters 
remains the same.

\section*{Appendix B: Li and Be mass fractions}

In this appendix we solve the evolution equation for the Li abundance in the 
outer convection zone with and without diffusion. The same calculation 
can be done to evaluate the evolution of the Be abundance.

The set of 
equations we wish to solve, (\ref{eq:mlidot1}) or (\ref{eq:xlidot2}), can be cast in the following general form: 
\begin{eqnarray}
&& \dot{X}^{\rm cz}_{\rm Li} = -\frac{ X^{\rm cz}_{\rm Li}}{\tau^{\rm cz}_{\rm pump}} + \frac{ X^{\rm cyl}_{\rm Li}}{\tau^{\rm cz}_{\rm pump}} - \frac{ X_{\rm Li}^{\rm cz} - X_{\rm Li,0} }{\tau_{\rm diff}} \mbox{  ,} \nonumber \\
\label{eq:xlidot3}
&& \dot{X}^{\rm cyl}_{\rm Li} =  \frac{ X^{\rm cz}_{\rm Li}}{\tau^{\rm cyl}_{\rm Li,pump}} - 2\frac{ X^{\rm cyl}_{\rm Li}}{\tau^{\rm cyl}_{\rm Li,pump}}\mbox{  .}
\end{eqnarray}
To reproduce equation (\ref{eq:mlidot1}) we set $\tau_{\rm diff} \rightarrow \infty$. Initial conditions are
\begin{equation}
\dot{X}^{\rm cz}_{\rm Li}(0) = \dot{X}^{\rm cyl}_{\rm Li} (0) = X_{\rm Li,0} \mbox{  .} 
\end{equation}

This set of equations has the solution
\begin{eqnarray}
&& X^{\rm cz}_{\rm Li}(t) = A e^{\lambda_1 t} + B e^{\lambda_2 t} + \frac{X_{\rm Li,0}}{\frac{\tau_{\rm diff}}{2\tau_{\rm pump}^{\rm cz}} + 1} \mbox{   ,}  \nonumber \\
&& X^{\rm cyl}_{\rm Li}(t) = \left(1 + \lambda_1 \tau_{\rm pump}^{\rm cz} + \frac{\tau_{\rm pump}^{\rm cz}}{\tau_{\rm diff}} \right) A e^{\lambda_1 t} + \left(1+\lambda_2 \tau_{\rm pump}^{\rm cz} + \frac{\tau_{\rm pump}^{\rm cz}}{\tau_{\rm diff}} \right) B e^{\lambda_2 t} + \frac{X_{\rm Li,0}}{\frac{\tau_{\rm diff}}{\tau_{\rm pump}^{\rm cz}}+2}\mbox{   ,} 
\end{eqnarray}
where 
\begin{equation}
\lambda_{1,2} = -\left(\frac{1}{\tau^{\rm cyl}_{\rm Li,pump}} + \frac{1}{2\tau_{\rm pump}^{\rm cz}} + \frac{1}{2\tau_{\rm diff}} \right) \pm \sqrt{ \frac{1}{4(\tau_{\rm pump}^{\rm cz})^2} + \frac{1}{2\tau_{\rm pump}^{\rm cz}\tau_{\rm diff}} +\left( \frac{1}{\tau^{\rm cyl}_{\rm Li,pump}} - \frac{1}{2\tau_{\rm diff}}\right)^2}\mbox{   .} 
\end{equation}  
Applying the initial conditions yields
\begin{eqnarray}
&& A = X_{\rm Li,0} \frac{\lambda_2}{\lambda_2 - \lambda_1} \frac{ 1 }{1  + 2\frac{\tau_{\rm pump}^{\rm cz}}{\tau_{\rm diff}}}  \mbox{   ,} \nonumber \\  
&& B = X_{\rm Li,0} \frac{\lambda_1}{\lambda_1 - \lambda_2} \frac{1}{ 1 + 2\frac{\tau_{\rm pump}^{\rm cz}}{\tau_{\rm diff}}}  \mbox{   .} 
\end{eqnarray}

Let us first examine the case with no diffusion ($\tau_{\rm diff}\rightarrow \infty$). In this case 
\begin{eqnarray}
&& x(t) = A e^{\lambda_1 t} + B e^{\lambda_2 t}  \nonumber \\
&& y(t) = \left(1 + \lambda_1 \tau_{\rm pump}^{\rm cz} \right) A e^{\lambda_1 t} + \left(1+\lambda_2 \tau_{\rm pump}^{\rm cz} \right) B e^{\lambda_2 t} \mbox{ ,}
\end{eqnarray}
with
\begin{equation}
\lambda_{1,2} = -\left(\frac{1}{\tau^{\rm cyl}_{\rm Li,pump}} + \frac{1}{2\tau_{\rm pump}^{\rm cz}} \right) \pm \sqrt{ \frac{1}{4(\tau_{\rm pump}^{\rm cz})^2} + \frac{1}{(\tau^{\rm cyl}_{\rm Li,pump})^2} } \mbox{  .}
\end{equation}  
In the limit where the mass of the outer convection zone becomes small compared with the mass
of the cylinder $C_{\rm Li}$, then $ \tau_{\rm pump}^{\rm cz} \rightarrow 0$. This implies that 
$X^{\rm cz}_{\rm Li}(t) \simeq X^{\rm cyl}_{\rm Li}(t)$, so that the respective concentrations of Li (or Be) in the 
convection zone and in the cylinder are the same. Moreover, they vary essentially as $e^{-t/\tau^{\rm cyl}_{\rm Li,pump}}$ since 
when $ \tau_{\rm pump}^{\rm cz}\ll \tau^{\rm cyl}_{\rm Li,pump} $, $\lambda_1 \simeq -1/\tau^{\rm cyl}_{\rm Li,pump}$ and 
$\lambda_2 \simeq -1/ \tau_{\rm pump}^{\rm cz}$.  The evolution on the timescale
$ \tau_{\rm pump}^{\rm cz}$ is extremely fast, leaving the longer timescale $\tau^{\rm cyl}_{\rm Li,pump} $ 
as the depletion timescale for the system. 
In the opposite limit (with the mass of the cylinder going to 0) 
it is easy to show that as before $X^{\rm cz}_{\rm Li}(t) \simeq X^{\rm cyl}_{\rm Li}(t)$ and the depletion timescale is now 
$\tau_{\rm pump}^{\rm cz}$. 

We expect that the effects of diffusion become important when the diffusion timescale 
into the convection zone $\tau_{\rm diff}$
becomes shorter than the advection timescale out of the convection zone $\tau_{\rm pump}^{\rm cz}$. 
In the limit $\tau_{\rm diff}/\tau_{\rm pump}^{\rm cz} \ll 1$ then the solution approximately becomes
\begin{eqnarray}
&& X^{\rm cz}_{\rm Li}(t) = X_{\rm Li,0} \left[  \frac{\tau_{\rm diff}}{2\tau_{\rm pump}^{\rm cz}} e^{- t/\tau^{\rm cyl}_{\rm Li,pump} } -  \frac{\tau_{\rm diff}^2}{2\tau_{\rm pump}^{\rm cz}\tau^{\rm cyl}_{\rm Li,pump} }  e^{-t/\tau_{\rm diff}} + 1 \right] \simeq  X_{\rm Li,0} \mbox{  ,} \nonumber \\
&& X^{\rm cyl}_{\rm Li}(t) = \frac{X_{\rm Li,0}  }{2} \left[  e^{-t/\tau^{\rm cyl}_{\rm Li,pump} }- \frac{\tau_{\rm diff}^2}{\tau_{\rm pump}^{\rm cz}\tau^{\rm cyl}_{\rm Li,pump} }  e^{- t/\tau_{\rm diff}} + 1 \right] \simeq \frac{X_{\rm Li,0} }{2}  \left( e^{-t/\tau^{\rm cyl}_{\rm Li,pump} } + 1 \right) \mbox{  ,} 
\end{eqnarray}
so that the Li abundance in the convection zone remains roughly primordial, as expected. The 
Li concentration in the cylinder gradually tends to half the primordial abundance, which is 
also expected since the cylinder itself becomes fully ventilated, and is a mixed region with equal incoming 
mass flux with primordial Li abundance, and incoming mass flux with zero Li abundance.

%


\end{document}